\documentclass{article}[11pt]
\pdfoutput=1 

\usepackage{graphicx}

\usepackage[dvipsnames,usenames]{color}

\usepackage{amsmath}                        

\definecolor{DarkGreen}{rgb}{0.5,0.8,0.6}   
\definecolor{RGBblack}{rgb}{0.0,0.0,0.0}    

\newcommand{\hei}[1]{\color{black}{#1} \color{black} }
\definecolor{grau}{rgb}{0.8,0.8,0.8}

\newcommand{\chen}[1]{\color{orange}}

\def\hei{\color{black}}

\newcommand{\jj}{\hei \rm}

\usepackage{setspace} 
\usepackage{url}
\usepackage{natbib}

\begin{document}

\title{On the Interval-Based Dose-Finding Designs}

\author{
        Yuan Ji\thanks{Correspondence: 1001 University Place, Evanston, IL USA, 60201. Email: koaeraser@gmail.com} \\
        {\small Program for Computational Genomics \& Medicine, NorthShore University HealthSystem} \\ 
        	{\small Department of Public Health Sciences, The University of Chicago}      
		\and     Shengjie Yang \\
         {\small Program for Computational Genomics \& Medicine,
           NorthShore University HealthSystem}}

\date{}
\maketitle

\begin{abstract}
\noindent The landscape of dose-finding designs for phase I clinical
trials is rapidly shifting in the recent years, noticeably marked by
the emergence of  interval-based designs. We categorize them as 
the iDesigns and
the IB-Designs. The iDesigns are originated by the toxicity
probability interval (TPI) designs and its two
modifications, the mTPI  and mTPI-2  designs. The IB-Designs started
as  the cumulative cohort design (CCD) and   is
 recently extended by the BOIN design. We
discuss the differences and similarities between these two classes of
interval-based designs, and compare their simulation performance with
popular   non-interval  designs, such as the CRM and 3+3
designs. We also show that in addition to the population-level
operating characteristics from simulated trials, investigators should
also assess the dose-finding decision tables from the implemented designs to
better understand the per-trial and per-patient behavior.  This is
particularly important for nonstatisticians to assess the designs with
transparency. We provide,
to our knowledge, the most comprehensive   simulation-based  comparative study on various  interval-based  dose-finding designs.

\end{abstract}

\section{Introduction} 
Phase I dose-finding trials aim to identify the maximum
tolerated dose (MTD), defined as the highest dose
with probability of toxicity close to or no higher than a targeted rate $p_T$ (say, $p_T
= 1/6$ or $1/3$). Classical dose-finding trials usually   prespecify  a set of $D$ ($D>1$) candidate dose levels, and each dose has an unknown probability $p_i$ of inducing dose-limiting toxicity (DLT) among patients. The trial starts from a low dose, treats a cohort of patients at the dose,  records their binary DLT  outcomes  up to a period of time, and decides the next dose level for treating future cohorts of patients. 
Iteratively, the trial   proceeds by sequentially treating future \jj
patients at a grid of 
predetermined  dose levels according to an appropriate design.  Usually additional safety rules are
implemented to safe-guard the enrolled patients (Goodman et al.,
1995), e.g., no skipping in dose escalation. Importantly, designs for
the phase I trials are confined by the following basic principle: 

\vskip 1.5em

\noindent {\bf Ethical Principle:} {\it 
One   should   not treat patients at doses perceived overly toxic, such as those believed to be above the MTD. 
}

\vskip 1.5em

Since 1989, the 3+3 design (Storer, 1989) and the CRM design (O"Quigley et al., 1990)  have been 
  the two main-stream statistical methods for phase I
dose-finding trials. The 3+3
design   has been   the most frequently applied approach in practice
and CRM the most  polished model-based methods in the
literature. CRM is based on a dose-response model that borrows
information across all the doses for decision making.  However, the main issues of CRM were lack of
standard implementation in practice and  challenges in interpreting the method to  clinicians. These issues have been improved
recently as many useful tools (e.g., the dfCRM and bcrm R packages) for CRM have
been developed. We refer to Cheung (2011) for an extensive discussion of CRM. 

Since 10 years ago, a new development in the design for dose-finding clinical
trials has been gradually transforming the research and practice. In
particular a new class of interval-based dose-finding designs have
been recognized, extended, and implemented by researchers and practitioners. In Cheung and Chappell
(2002), the authors introduced the concept of indifference interval in
defining the MTD, and Ji et al. (2007ab) first introduced a design that
solely depends on properties of toxicity probability
intervals (TPI). That is, the statistical inference for dose-finding
process is dependent on the posterior probability of   the TPIs,
instead of a point estimate, that describes the toxicity of a dose. Ji
et al. (2010), Ji and Wang (2013) and Yang et al. (2015) further developed the modified TPI (mTPI) designs to improve the performance of TPI. These methods share the same
idea of assessing the posterior probabilities that the toxicity rate $p_i$ of a
dose $i$ falls into three intervals $(0, p_T - \epsilon_1)$, $(p_T - \epsilon_1, p_T
+ \epsilon_2)$, and $(p_T + \epsilon_2, 1)$,   represented as $Pr\{p_i \in (p_T - \epsilon_1, p_T + \epsilon_2) \mid data\}$ for example,  and conducting up-and-down
type of dose-finding decisions
using the posterior probabilities of the three intervals. Here
  $\epsilon_{1} $ and $\epsilon_2$ \jj are small fractions that reflect investigators'
desire about how accurate they want the MTD to be around $p_T$, the
target value. Usually $\epsilon_{1,2}   \le  0.05.$ For example, when $p_T
= 0.3$ and $\epsilon_{1,2} = 0.05$, the three intervals become $(0,
0.25)$, $(0.25, 0.35)$, and $(0.35, 1)$. This means that doses with
toxicity probabilities between 0.25 and 0.35 can be considered as the
MTDs, and those lower than 0.25 or higher than 0.35 are too low or too
high relative to the target value $p_T=0.3$. Based on this simple
logic, mTPI calculates the posterior probabilities of
these three intervals for a given dose and use a decision
theoretic framework to guide the optimal dose finding in mTPI. The
authors show that the decisions of mTPI are optimal, minimizing a
posterior expected risk. Recently, Guo et al. (2017) proposed an
extended version called mTPI-2, which is based on the same interval
decisions but with finer intervals to mitigate some practical concerns
in mTPI. 

During  the same time period since 2007, an alternative class of interval designs
was developed, represented by the cumulative cohort design (CCD,
Ivanova et al., 2007)
  and more recently an extended version as  the Bayesian optimal interval design (BOIN, Liu and Yuan,
2015). In CCD, the authors propose to calculate   a point estimate  $\hat{p}_i = x_i /n_i$
where $x_i$ is the number of patients experiencing the dose limiting
toxicity (DLT) events and $n_i$ is the number of patients that have
been treated at  current  dose $i$. Then CCD compares $\hat{p}_i$ to
three intervals of the same format as in mTPI. If $\hat{p}_i$ falls in
the lower   interval $(0, \phi_1),$ middle interval $(\phi_1, \phi_2)$, or upper interval $(\phi_2, 1)$,  the decision is to escalate to  dose 
$(i+1)$, stay at dose $i$, or de-escalate to dose $(i-1)$,
respectively. The BOIN design uses the same type of decision rules
except that it re-calibrates the middle interval of    $(\phi_1, \phi_2)$  to a
new interval $(\lambda_{e}, \lambda_{d})$ which typically satisfies
$\phi_1 < \lambda_e < \lambda_d < \phi_2$. The
authors   argue that the new boundaries $\lambda_{e}$ and $\lambda_d$ are optimal as they minimize an  error function. 

We will review these two developments of interval-based designs and
demonstrate the key similarities and differences among them. 
In Section 2, we review TPI, mTPI, and mTPI-2 as the   ``iDesigns",   with ``i" standing for ``intervals",   which 
use probabilities of the toxicity probability intervals in a Bayesian
framework. We also review CCD and BOIN   as the ``IB-Designs",   with
"IB" standing for "interval boundaries". ,   which  use the point estimate
$\hat{p}_i$ and fixed interval boundaries  for inference.  
 In addition, we delineate an important difference between the interval-based designs and the CRM  design.  
  In Section 3 we report operating characteristics of different designs based on objective, crowd-sourcing, and large-scale simulation studies. In Section 4 we propose a new  criterion  for the evaluation of different designs other than the operating characteristics from simulated trials.  We end
with a discussion in Section 5.

\section{  Review of Designs } \label{sec:review}
\subsection{The iDesigns} \label{sec:iDesign}
We first review a class of designs based on probability of toxicity probability intervals (TPIs). We call these designs the iDesigns.

\paragraph*{The TPI Design} 
  Consider three specific TPIs,  the under-dosing interval (UI) $(0, p_T -
\epsilon_1)$, the equivalence interval (EI) $(p_T - \epsilon_1, p_T +
\epsilon_2)$, and the over-dosing interval (OI) $(p_T + \epsilon_2,
1)$. Suppose the toxicity rate of a dose $i$ is $p_i$. When
$p_i$ falls into UI, EI, or OI, dose $i$ is considered below,
equivalent to, or above the MTD, respectively. If dose $i$ is currently used for
  treating   patients, then the decision  is to ``escalate'' (E) to dose
$(i+1)$, ``stay'' (S) at dose $i$, or ``de-escalate'' (D) to dose
$(i-1)$, respectively. In other words, {\bf decision E is associated with
UI, S associated with EI, and D associated with OI.} This mimics a typical up-and-down type of decision
in dose finding (see Stylianou and Flournoy, 2002  for a discussion). Therefore, as long as one can
accurately infer to which interval $p_i$ belongs, one can easily
decide the dose for future patients in the trial.

 In the TPI design,  Ji
et al., (2007a) use $(0, p_T - K_1 \sigma_i)$, $(p_T - K_1 \sigma_i, p_T + K_2 \sigma_i)$, and $(p_T+ K_2 \sigma_i, 1)$ to describe the
EI, UI, and OI, respectively, where $K_{1,2}$ are
two prespecified constants and $\sigma_i$ is the posterior standard
deviation of $p_i$. They define a penalty (loss) function that maps 
 decisions  E, S, or D and the value of $p_i$ to a value of loss. For
example, if $p_i$ is in UI and the decision is $E$, there is no
loss. However, if $p_i$ is in UI and the decision is $D$ or $S$, there
is a positive loss. The positive loss values quantify the violation of the
underlying ethical constraints   that patients must not be treated at doses above the MTD.   Under this decision theoretic framework, the
authors derived the Bayes rule which chooses one of the three
decisions (E, S, D) that minimizes the posterior expected loss. For
statistical inference, the TPI design models the observed binary
toxicity outcomes use a binomial likelihood, and models $p_i$ with
independent beta prior distributions. This anti-intuitive choice of
prior models have been show to work well with up-and-down rules, and
are in sharp contrast to other model-based designs, such as the CRM
design. However, it appears the use of independent beta
priors is being increasingly adopted by various new methods, including
the most recent development of a semi-parametric CRM design (Clertant
and O'Quigley, 2017).

\paragraph*{The mTPI Design} In the 
mTPI design (Ji et al, 2010), the authors made $\epsilon_{1}$ and $\epsilon_2$ two
user-specified constants, with the notion that the equivalence
interval $(p_T - \epsilon_1, p_T + \epsilon_2)$ should be provided
by the trial clinicians.  This is similar to the concept of
user-provided effect sizes in designing randomized clinical trials. In particular, $(p_T - \epsilon_1)$ should be
the lowest toxicity probability that the clinician would feel
comfortable not to escalate, and $(p_T + \epsilon_2)$ should be the
highest toxicity probability that the clinician would feel comfortable
not to de-escalate. Therefore, any   dose with  probability in between $(p_T -
\epsilon_1, p_T + \epsilon_2)$ can be considered as the
MTD. Under the same decision theoretic framework, mTPI derives the
optimal rule (Bayes' rule) that minimizes the posterior expected
loss.   The optimal rule is   simple and  is \jj equivalent to 1) identifying the
interval among UI, EI and OI that has the largest unit probability
mass (UPM, defined below) and 2) choosing the
decision E, S, or D that is associated with the interval. Here, UPM of an interval is
defined as the ratio of the posterior probability $p_i$   belonging   to the
interval and the length of the interval. For example, 
\begin{equation}
\mbox{UPM(EI)} =
\frac{Pr\{p_i \in (p_T - \epsilon_1, p_T + \epsilon_2) \mid data\}}{\epsilon_1+\epsilon_2}. \label{eq:UPM}
\end{equation}

\paragraph*{The mTPI-2 Design} Although mTPI uses the Bayes' rule -- an optimal decision rule that
minimizes the posterior expected loss  to guide
decision making, the statistically optimal
decisions may be at odds with the {\bf Ethical Principle}   stated in Section 1 for
clinical  trials. For example, suppose at the current dose 3 out of 6
patients experienced DLT and the trial targets probability
$p_T=0.3$ for the true
MTD; under   the Bayes' rule,  the mTPI
design would instruct to ``S'', stay at the current dose. Since the empirical rate is 3/6=0.5
and the target rate is 0.3, oftentimes   the dose would be considered
above the MTD and therefore the desirable decision is   ``D'',
de-escalate. However, there is not clear guidelines in the community
on what decisions are acceptable. And even though there is a large
uncertainty associated with the data of only 6 patients, many safety
review boards would express concerns if a design chooses ``S'' when
the empirical data is 3/6.  Interestingly, when the trial data shows 4   out  of 8 or 5 out
   of  10 patients experienced DLT, even though the empirical rate is
still 0.5, mTPI would instruct to ``D'',
de-escalate to the lower dose.  Therefore, the mTPI design automatically accounts for the variability of the data in the decision making, which is considered as a key principle in statistics inference.  

However,   in trial implementation, patient safety should be strictly
protected. To this end, Guo et al. (2017) conduct a detailed investigation. 
They note that just as 
model selection criteria like   AIC (Akaike, 1974) or BIC (Schwarz, 1978)  include a
penalty for model size and a score (log-likelihood) for model fitting,
the Bayesian  decision theoretic  framework in mTPI intrinsically treats the
three intervals as three models, and penalize models   based on  the model size which is
the length of each interval.   As a result, the Bayesian Occam's razor principle favoring parsimonious models (Jefferys and Berger, 1992) is seen in mTPI.  
 Guo et
al. (2017) propose a simple 
remedy known as the mTPI-2
design, in which the UI and OI are broken into smaller subintervals having the same length as the EI. Then the interval having 
the largest UPM value will be associated with the corresponding dose-finding
decision. For example, suppose $p_T=0.3$ and the EI is $(0.25, 0.35)$;
mTPI-2 then breaks the OI $(0.35, 1)$ into subintervals $(0.35, 0.45),
(0.45, 0.55), $ and so on. If any subinterval in the OI has the
largest UPM,  the decision is ``D'', to de-escalate; if any
subinterval in the UI has the largest UPM,  the decision is ``E'',
to escalate; and if the EI has the largest UPM, the decision is
``S'', to stay. Guo et al. (2017) show that this again is an optimal
decision that minimizes the posterior expected loss for a 0-1 loss
function. In addition, they demonstrate that the new Bayesian inference is
now not affected by the Occam's razor, as all the subintervals (except
for the two boundary ones) have the same length. Consequently, when
$p_T=0.3$ and 3 out of 6 patients experienced DLT, mTPI-2's
decision is now ``S'', to stay.

\subsection{The IB-Designs}\label{sec:ibdesign}
Almost parallel to the development of the TPI, mTPI, and mTPI-2 methods,
a different class of interval-based designs has been developed and
extended. These methods are based on predetermined decision rules that can be summarized in three steps: 
\begin{enumerate}
  \item Obtain a point estimate of $p_i$, say  $\hat{p}_i = x_i/n_i$.
  \item Decide three toxicity probability intervals   $(0, \phi_1), (\phi_1, \phi_2), (\phi_2, 1)$  so that $\phi_1 < p_T < \phi_2$ are  two boundaries that define the equivalence interval $(\phi_1, \phi_2)$ (they are similar to $\epsilon_{1,2}$ in the mTPI and mTPI-2 designs),
  \item Escalate, stay, or de-escalate if $\hat{p}_i$ falls into the three intervals, respectively. 
\end{enumerate}
As can be seen,   these rules use intervals differently from the
iDesigns.   The iDesigns use $Pr(p_i \in \mbox{an interval} \mid
data)$ for decision making, while here the decisions are based on
comparison of $\hat{p}_i$ and intervals. In other words,   here the intervals serve as boundaries for assessing the point estimate $\hat{p}_i$. Therefore, we call these interval-based designs the IB-Designs where ``IB" stands for interval boundaries.

\paragraph*{The CCD Design}  Under the CCD design (Ivanova et al.,
2007), if $\hat{p}_i$
falls into the lower interval $(0, \phi_1)$, escalate to dose $(i+1)$; the middle
interval $(\phi_1, \phi_2)$, stay at dose $i$; the upper interval $(\phi_2, 1)$, de-escalate to dose
$(i-1)$.  Because only change of one dose level is allowed in
escalation or de-escalation, the CCD design generates a Markov chain
of toxicity outcomes, and the authors prove asymptotic convergence
of the design.   In the CCD design,   the authors recommend a simpler form by expressing  the lower cutoff
toxicity probability $\phi_1 = (p_T - \Delta)$ and the upper cutoff
$\phi_2 = (p_T + \Delta),$ which have equal distance to $p_T$. Then
the above 1-3 steps are applied to guide the dose finding trial.  
Choices of $\Delta$ are given by Ivanova et al. (2007) for moderate
sample size ($\le 20$) and six dose levels. We list the
recommended values below: 1) $\Delta = 0.09$ for $p_T \in \{0.10,
0.15, 0.20, 0.25\}$; 2) $\Delta = 0.10$ for $p_T \in \{0.30, 0.35\}$;
3) $\Delta = 0.12$ for $p_T = 0.40$; and 4) $\Delta=0.13$ for $p_T \in
\{0.45, 0.50\}$. 

\paragraph*{The BOIN Design} Liu and Ying (2015) extend CCD and develop the BOIN designs, with local and global BOIN as two
versions. They state that BOIN is an improvement of CCD
since it uses interval boundaries that are optimal and vary with
enrollment counts and dose levels. In BOIN, the dose-finding problem is cast slightly differently. Now
consider $0<\phi_1 <p_T < \phi_2 <1$ two user-provided values (instead of theoretically derived in CCD) that represent
the lower and upper bound of the equivalence interval for $p_T$. In this sense, $\phi_1$ and $\phi_2$ are given and decided by the trial investigators, which is similar to the mTPI and mTPI-2 designs. Therefore, a dose with a toxicity probability falling into $(\phi_1, \phi_2)$
can be considered as an MTD candidate.   Suppose dose $i$ is currently used to treat patients in the trial.  In the local BION design, an
ad-hoc optimization procedure changes the two values of  $\phi_1$ and
$\phi_2$ to two new values called $\lambda_e(i)$ and $\lambda_d(i)$ so that
the interval $(\lambda_e(i), \lambda_d(i))$ is nested in $(\phi_1,
\phi_2)$, i.e.,  $\phi_1 < \lambda_e(i) < \lambda_d(i) <
\phi_2$.   Interestingly, the optimal values of $\lambda_e(i)$ and $\lambda_d(i)$ do not depend on dose level $i$ based on the local BOIN design. Therefore, we drop index $i$   in $\lambda_e(i)$ and $\lambda_d(i)$  hereafter for local BOIN. The local BOIN design first examines if  $\hat{p}_i$ falls into
one of the three intervals $(0, \lambda_e]$, $(\lambda_e, \lambda_d)$,
$[\lambda_d, 1)$, and  escalates to dose $(i+1)$, stays at dose $i$,
or de-escalates to dose $(i-1)$, accordingly. In other words, the local
BOIN design uses the same concept for dose escalation as the CCD
design, except BOIN changes the original user-provided boundary
$\phi_{1,2}$ to $\lambda_{e,d}$ based on an optimization
criterion. Liu and Ying (2015)    also develop a global BOIN design,
in which the two boundary values depend on the number of treated
patients at each dose level $i$.   That is, $\lambda_{e,d}$ in
global BOIN do depend on dose level $i$.  Although conceptually the global BOIN design is more advanced, the authors conclude that the global BOIN design does  not perform as
well as the local BOIN design in their simulations. Therefore, the
authors recommend the local BOIN design for practical
implementation, which is quite similar to the CCD design since the
boundaries of both designs are not dependent on dose levels.   We
refer to BOIN as the local BOIN design hereafter.  

\subsection{The iDesigns and IB-Designs}\label{two iDesigns}
 Both iDesigns and IB-Designs   use an equivalence interval $(a, b),$  $a < p_T < b,$ to
convey the fundamental concept of the interval designs -- {\bf to represent
the MTD with an interval instead of a single value.} BOIN and CCD  both use an equivalence interval $(\phi_1, \phi_2)$ as the boundaries for $\hat{p}_i$.  While CCD uses theoretical derivation to determine $\phi_{1,2}$, BOIN requires users provide $\phi_{1,2}$ first and then derives a new set of boundaries $\lambda_{e,d}$ based on $\phi_{1,2}$ as the equivalence interval.  
 Table \ref{tab:designs} lists the main common and different features
 of the designs. The main difference between iDesigns and IB-Designs
 is that iDesigns use the posterior probability that $p_i$ falls into
 the intervals as the foundation for statistical inference and
 decision making, while the IB-Designs relies on the point estimate
 $\hat{p}_i$ and directly compares it with the interval  boundaries.  Also, the
 IB-designs do not use a   classical decision-theoretic framework
 (Berger, 2013) that involves loss, risk, and decision rules as the
 essential elements. The optimality suggested by BOIN is based on an ``error'' function $\alpha(\lambda_e, \lambda_d)$ (Liu and Ying,
2015) to mimic the type-I error in hypothesis testing. In contrast,
the iDesigns use a formal decision theoretic setup, with a
loss/penalty function, and aims to find the optimal decision rule that
takes D, S, or E as {\it actions} and minimizes the posterior expected loss. 

Ying et
al. (2016)  argue that the use of UPM in mTPI is not desirable since it produces
questionable decisions such as ``S'' when 3 out of 6 patients
experienced toxicity at a given dose. However, Guo et al. (2017)
show that these decisions in mTPI are theoretically justified under the principle
of Occam's razor. But because of the {\bf Ethical Principle} in
phase I   clinical   trials, the   mTPI  decisions might not be perceived as
ethically sound in practice. Therefore, Guo et al. (2017) propose the mTPI-2 design which blunts the Occam's razor in order to  achieve
theoretical optimality and ethical satisfaction at the same
time. Later  in this manuscript,  we show that mTPI-2 and BOIN perform quite similarly. 

\begin{table}
\resizebox{1\textwidth}{!}{
\begin{tabular}{|l|c|c|c|c|c|}\hline
 Criterion & \multicolumn{3}{|c|}{The iDesigns} & \multicolumn{2}{|c|}{The IB-Design} \\ \cline{2-6} 
        & TPI (2007) & mTPI (2010)  & mTPI-2 (2016) & CCD (2007) &BOIN (2015) \\ \hline
1. $p_T \pm constant$ as equivalence interval & Y & Y & Y & Y & Y \\
2. Loss/penalty function & Y & Y & Y & \fbox{N}  & \fbox{N} \\
3. Safety rule $Pr(p_i > p_T | data) < 0.95$ & Y & Y & Y & \fbox{N} & Y \\
4. Isotonic$^*$ for MTD selection & Y & Y & Y & Y & Y \\
5. A decision table of (D, S, E) & Y & Y & Y & Y & Y \\
6. Inference using probabilities of the intervals & Y & Y & Y & \fbox{N} & \fbox{N}\\
7. Additional safety rule and dose exclusion rule & Y & Y & Y & \fbox{N} & Y\\
\hline
\end{tabular}
}
\caption{Summary of different features between the three iDesigns, TPI, mTPI, mTPI-2, and two IB-Designs, CCD and BOIN. "Y" or "N" represents the feature is or  is  not present in the design.   $^*$: All the methods use isotonic transformation to select the MTD at the end of the trial.  }\label{tab:designs} 
\end{table}

  There are a few other iDesigns and IB-Designs in the
literature  with similar features as the ones we reviewed so
far. For example, there is a frequentist-version of the mTPI design,
called TEQR (Blanchard and Longmate, 2011) as an iDesign. The Group design
(Gezmu and Flournoy, 2006) that precedes the CCD design belongs to the
category of IB-Designs.  TEQR uses the same concept as the mTPI but
resorts to a frequentist inference. The Group design decides the next
dose assignment only using the outcomes of the current cohort of
subjects   at a dose level.  We omit reviewing these methods in
details as they largely overlap with the iDesigns and IB-Designs that
have been discussed. We also omit the detailed review of Cheung and
Chappell (2002) who pioneered the use of indifference interval
boundaries in their work, as their work is largely aimed at the
sensitivity of the CRM design, a non-interval dose-finding design.


\subsection{A Key Difference to CRM}\label{sec:compCRM}
The iDesigns and IB-Designs  differ from the CRM design in one
important aspect. The interval designs use up-and-down 
and fixed decision rules when patient outcomes are observed at a given dose.   For example,  suppose dose $i$ is the current dose and for a given set of observations $(x_i, n_i)$
at the current dose $i$, the next dose level for treating future patients under the interval designs
is fixed, regardless of 
data on other doses and the dose level $i$. CRM is different and uses random
rules. That is, given $(x_i, n_i)$ at dose $i$, the decision of
escalation, de-escalation, or staying can vary and depends on $(x_j, n_j)$ on other
doses $j$'s. That is, CRM ``borrows information'' from all the doses,
which is  a sound statistical principle. However, as seen in
 mTPI,  for dose-finding trials with the {\bf Ethical Principle}  and small
sample size, sound statistical principles do not necessarily lead to
sound ethical decisions. For example, the original CRM method
would give ethically unacceptable decisions, which led to 
many additional ethical rules later on   (Goodman et al., 1995;
Babb et al., 1998).  In contrast, the interval-based designs do not formally borrow information across doses in the statistical estimation. However, because they use up-and-down rules (D, S, and E), there is intrinsic borrowing information   in the Markov-dependent manner (Ivanova et al., 2007)  across different cohorts of patients. To see this, let the current dose be dose level $i$. Then under iDesigns and IB-Designs, dose $(i-1)$ must have been perceived as below the MTD and dose $(i+1)$ is either un-used or must have been perceived as above the MTD. 

\section{Numerical Examples}\label{sec:example}
\subsection{Designs Under Comparison}
We conducted three extensive simulation studies to assess the performance
of four different types of designs, 1) 3+3 as an algorithmic design,
 2) CRM as the classical
non-interval-based design, 3) mTPI and mTPI-2 as  iDesigns, and 4) BOIN as an IB-Design. For
BOIN, we implemented three versions of the local BOIN design to
examine the sensitivity of the design to different specifications of
the equivalence boundaries. Recall that  BOIN uses the simple rule that 
compares $\hat{p}_i = x_i / n_i$ with two boundaries, $\lambda_e$ and $\lambda_d$. If
$\hat{p}_i$ is in $(0, \lambda_e]$, $(\lambda_e, \lambda_d)$, or
$[\lambda_d, 1)$, the decision is E, S, or D, respectively. The two
$\lambda$'s are calculated as a linear and one-to-one functions of two user-provided values $\phi_1$ and $\phi_2$ where $0<\phi_1 < \lambda_e < \lambda_d < \phi_2 <
1$. The values of $\phi_1$ and $\phi_2$ are   determined  so that doses with toxicity probabilities smaller than $\phi_1$ or larger
than $\phi_2$ will not be considered as MTD. In order to compare BOIN with
mTPI and mTPI-2 under different BOIN assumptions, we 
implemented three versions of BOIN. In the first version we used the default choice
in which Liu and Ying (2015) suggested setting $\phi_1 = 0.6 p_T$ and
$\phi_2 = 1.4 p_T$ regardless of trials, diseases and dose levels. We call this version $\mbox{BOIN}_{\mbox{default}}.$ The second version is to set $(p_T - \phi_1) =
\epsilon_1$ and $(\phi_2 - p_T) = \epsilon_2$, so that the user-provided
equivalence interval is the same  for BOIN and the iDesigns. We call the
second version $\mbox{BOIN}_{\mbox{epsilon}}.$ However, as can
be seen above BOIN actually uses  $\lambda_{e}$  and $\lambda_{d}$, not the user-provided $\phi_{1,2}$ for
decision making. For
this reason, we tried the third version of BOIN, called $\mbox{BOIN}_{\mbox{lambda}},$ which sets $(p_T - \lambda_e) =
\epsilon_1$ and $(\lambda_d - p_T) = \epsilon_2$. This way, the actual intervals for decision making between BOIN and iDesigns are identical.  For example, if $p_T = 0.3$, the mTPI and mTPI-2 designs would
by default assume $(0.25, 0.35)$ as the equivalence interval,
corresponding to $\epsilon_1 = \epsilon_2 =0.05.$ In contrast, BOIN
would have three versions: 
$\mbox{BOIN}_{\mbox{default}}$ assumes $\phi_1 = 0.18$ and
$\phi_2 = 0.42$, which gives 
$\lambda_e = 0.236$
and   $\lambda_d = 0.358.$  Therefore, $\mbox{BOIN}_{\mbox{default}}$ escalates if the current dose
$i$ has
$\hat{p}_i < 0.236$, de-escalates if $\hat{p}_i > 0.358$, or stay if $0.236 < \hat{p}_i <
0.358.$ For $\mbox{BOIN}_{\mbox{epsilon}},$ it has $\phi_1=0.25$ and $\phi_2=0.35$ which gives 
$\lambda_e=0.275$ and $\lambda_d = 0.325$. And for
$\mbox{BOIN}_{\mbox{lambda}}$, one would use $\lambda_e=0.25$ and
$\lambda_d = 0.35$ directly.   Note this implies that $\phi_1 = 0.205$ and $\phi_2 = 0.402.$  

\subsection{Simulation Scenarios}\label{sec:scenario}
We considered   three  sets of scenarios 
to comprehensively compare the designs. 

\begin{figure}[htbp]
\begin{center}
\includegraphics[scale=0.35]{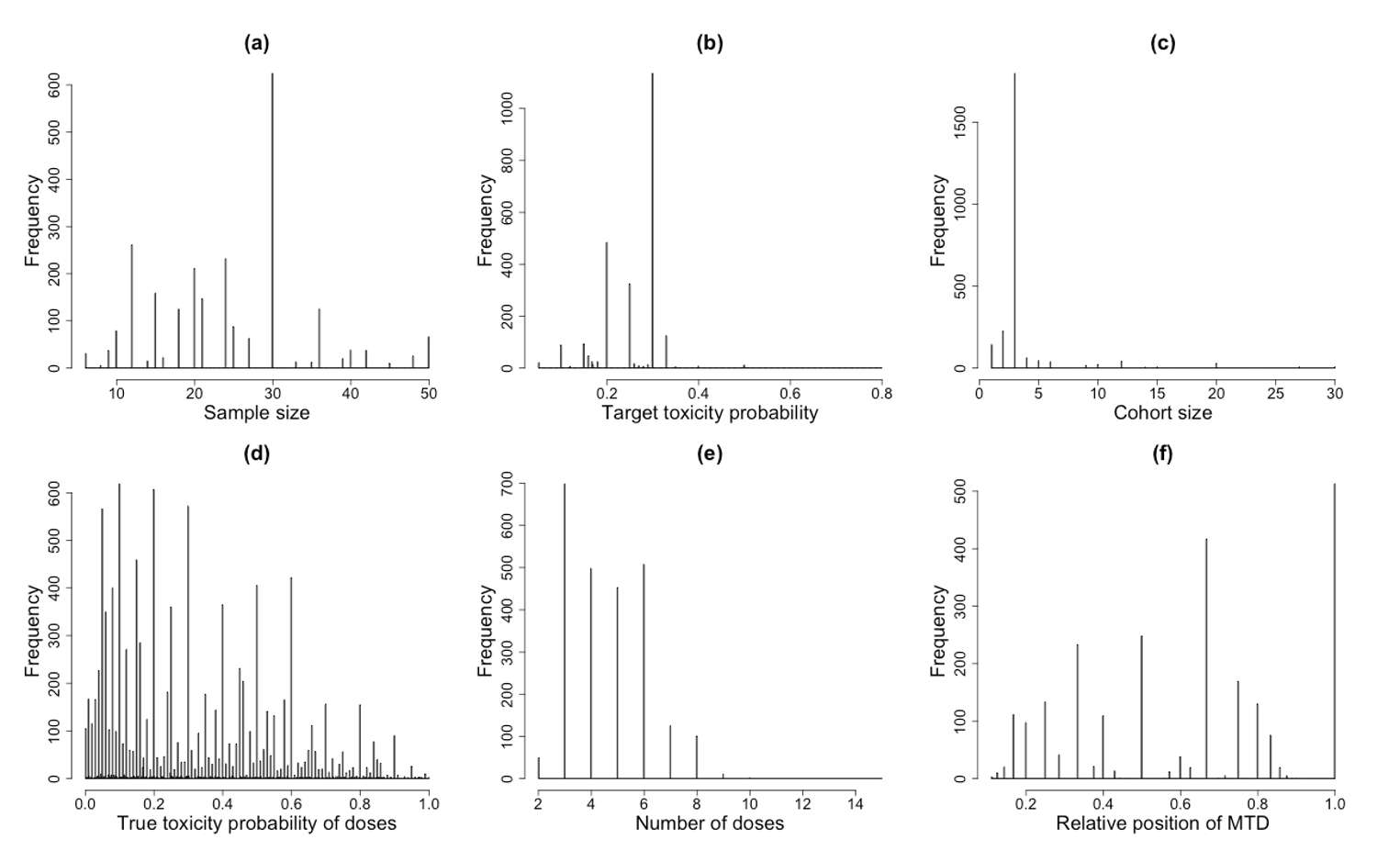}
\caption{Summary of 2,447 scenarios from the NGDF users. Panels
  (a)-(e) are the sample size, target toxicity probability $p_T$,
  cohort size of the trial, the true toxicity probability of the
  doses in all the scenarios, and the number of doses in all the
  scenarios, respectively. Panel (f) gives the relative position of
  each dose in each scenario. Suppose a scenario has $d$ doses, and
  each dose is indexed by $i$, $i=1, \ldots, d$. Then the relative
  position of dose $i$ is $i/d$. }\label{fig:ngdf-sc}
\end{center}
\end{figure}

\paragraph{Set one.} The first set of scenarios was
generated via ``crowd-sourcing'' (Yang et al., 2015). Specifically, we
recorded the scenarios generated by the real users of the online tool
 Next-Gen DF (NGDF,   Yang et al., 2015),  which provided web-based software for simulating
phase I dose-finding trials using mTPI-2, mTPI, CRM, and 3+3. We
collected 2,447 scenarios generated by NGDF users from April 16, 2014 to
September 8, 2016, which are unique in sample size, target toxicity rate $p_T$,
cohort size, and true toxicity probability. Figure \ref{fig:ngdf-sc}
summarizes the scenarios using six different characteristics. It can
be seen the scenarios are quite different in terms of these
characteristics. 

\paragraph{Set two.} We adopted the 42 scenarios in Ji and Wang (2013), which provides 14 scenarios for each of the three $p_T$ values, 0.1, 0.2, and 0.3. Although the number of scenarios is small, they are representative of different dose-response shapes and do not over- or under-represent any particular shapes. See Appendix B for details. 

\paragraph{Set three.}  Following Paoletti et al. (2004), we generated
1,000 scenarios based on a model-based approach. That is, we use an
ad-hoc probability model (see below) to generate the scenarios.

\begin{figure}[htbp]
\centering
\includegraphics[scale=0.36]{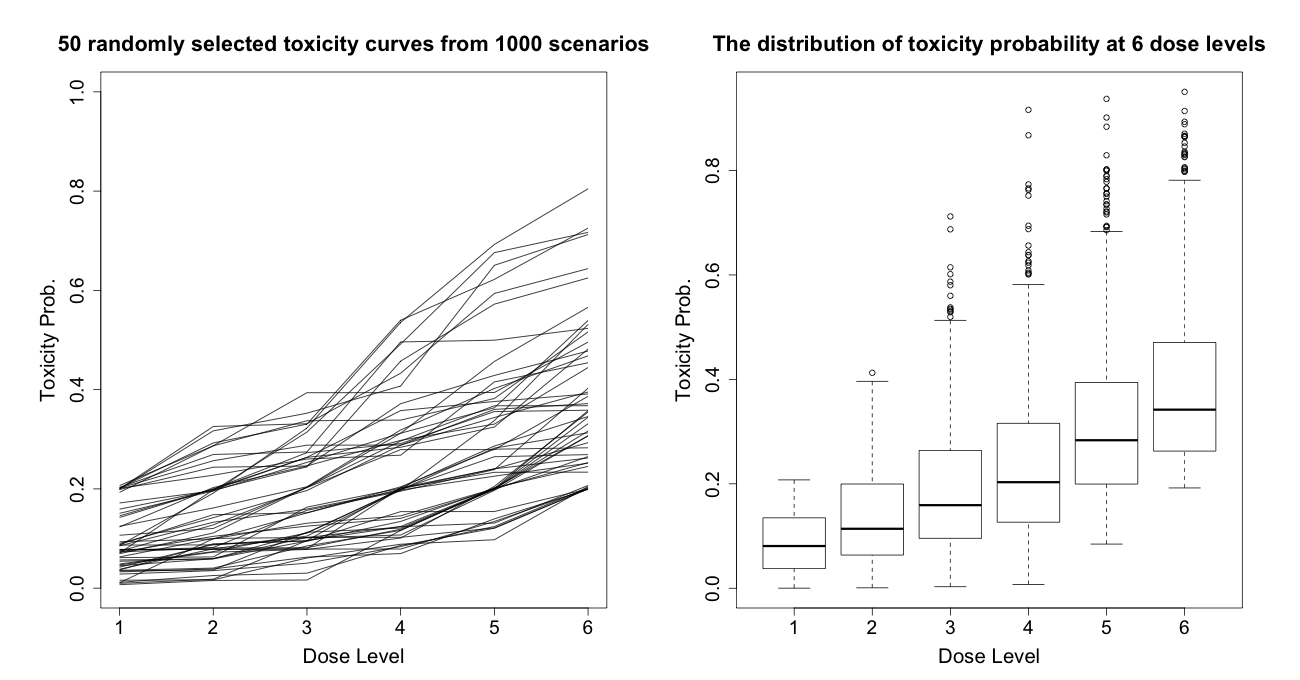}\\
\caption{Summary of the 1,000 scenarios with $p_T=0.2$ generated from the random scheme in Paoletti et al. (2004). Left: a randomly selected 50 scenarios. Right: the barplots of the true toxicity probability for each of the six dose levels across 1,000 generated scenarios.}\label{fig:1000sc}s
\end{figure}

\begin{description}
  \item[\textbullet] Assume that there are $d$ doses in the scenario, $d>1.$ Fix
    the target toxicity probability value
    $p_T$, say $p_T=0.2.$
   \item[\textbullet] With equal probabilities select one of the $d$
                  doses as the MTD. Without loss of generality, denote
                  this selected dose level by $i$.
   \item[\textbullet] Generate a random value $p_i = \Phi(\xi_i)$ where $\Phi(\cdot)$ is
     the cumulative density function (CDF) of the standard normal
     distribution, and assign
     $p_i$ to be the toxicity rate of the MTD for this scenario. Here
     $\xi_i\sim N(\Phi^{-1}(p_T),0.01^{2})$, where $\Phi^{-1}(\cdot)$
     denotes the inverse CDF of the standard normal distribution. The
     term $\xi_i$ perturbs the toxicity probability of the MTD so
     that it is not identical to $p_T$.
  \item[\textbullet] Generate the toxicity rates of the two adjacent doses,
        $p_{i-1}$ and $p_{i+1}$ by 
$$
p_{i-1} =
\Phi
\left[\Phi^{-1}(p_i)-\left\{\Phi^{-1}(p_i)-\Phi^{-1}(2p_T-p_i)\right\}\times
  I\{\Phi^{-1}(p_i)>\Phi^{-1}(p_T)\}-\xi_{i-1}^2
  \right]
$$ 

and

$$p_{i+1} =
\Phi\left[\Phi^{-1}(p_i)+\left\{\Phi^{-1}(2 p_T-p_i)-\Phi^{-1}(p_i)\right\}\times
  I\{\Phi^{-1}(p_i)<\Phi^{-1}(p_T)\}+\xi_{i    + 1}^2\right]$$ 
where $\xi_{i-1}\sim N(\Phi^{-1}(\mu_1),0.1^{2})$, $\xi_{i+1}\sim N(\Phi^{-1}(\mu_2),0.1^{2})$, and $I(\cdot)$ is an indicator function. 

		\item[\textbullet] Generate the toxicity probabilities iteratively
                  for the remaining dose levels according to $p_{i-2}=\Phi[\Phi^{-1}(p_{i-1})-\xi_{i-2}^2]$ and $p_{i+2}=\Phi[\Phi^{-1}(p_{i+1})+\xi_{i+2}^2]$, and $p_{i-3}=\Phi[\Phi^{-1}(p_{i-2})-\xi_{i-3}^2]$ and $p_{i+3}=\Phi[\Phi^{-1}(p_{i+2})+\xi_{i+3}^2]$, and so on, where $\xi_{i-2},\xi_{i-3},\ldots,\sim N(\Phi^{-1}(\mu_3),0.25^{2})$ and $\xi_{i+2},\xi_{i+3},\ldots,\sim N(\Phi^{-1}(\mu_4),0.25^{2})$.

 \end{description}
We set the values of $\mu$'s 
such that the average distance at each side of MTD
is less than 0.1. This generates a set of ``smooth" scenarios where
toxicity gradually increase along the dose levels.  See Figure \ref{fig:1000sc} for details.  


\subsection{Results}
\paragraph{Definition of true MTD} In the interval designs, because
the doses with probability of toxicity falls into the equivalence
interval $(p_T-\varepsilon_1, p_T+\varepsilon_2)$ are acceptable MTDs, we provide a clear definition of true
MTDs for a given scenario. 
\begin{enumerate}
	\item If $(p_T-\varepsilon_1) \leq p_i \leq (p_T+\varepsilon_2)$, then dose $i$ is the true MTD. If more than one dose satisfies the condition, then all of them are considered the true MTDs.

	\item If there is no $p_i$ meeting condition 1 above, the true MTD is the maximum dose level $i$ of which the true toxicity probability $p_i < p_T$.
        
	\item If the MTD could not be identified (e.g., if all the doses have toxicity probabilities $> p_T$), the correct decision is not to select any dose and the true MTD is set as `none'. In other words, selecting any dose as the MTD would be considered as a mistake.
\end{enumerate}

\paragraph{Results for set one.} We implemented mTPI-2, mTPI, all three BOIN versions, CRM and 3+3  based on the 2,447 crowd-sourcing scenarios.  For each scenario, 2,000 simulated trials were conducted on computer.  We compared designs' operating characters in terms of two simple metrics, {\it safety} and {\it reliability} (Ji and Wang, 2013). These two metrics capture the most important and fundamental properties of a dose-finding design: patient safety and ability to identify the true MTD. 
 {\it Safety} is the average percentage of the patients treated at or
 below the true MTD across the simulated trials for a given scenario,
 and {\it reliability} is the percentage of simulated trials selecting
 the true MTD for a given scenario. In some literature, {\it
   reliability} is also called PCS, standing for the percentage of
 correct selection of the true MTD.  Usually there is a tradeoff between the two metrics. That is, a design with better safety is usually associated with less reliability and vice versa.  This is because in order to correctly identify the true MTD, a design must accurately infer doses both below and above the MTD, which means assigning some patients to doses above the MTD to learn their   high  toxicity probabilities. Doing that would lower the {\it safety} metric of the design.

\begin{figure}[htbp]
	\begin{center}
	\resizebox{0.9\textwidth}{3.5in}{
	\begin{tabular}{c}
	\includegraphics{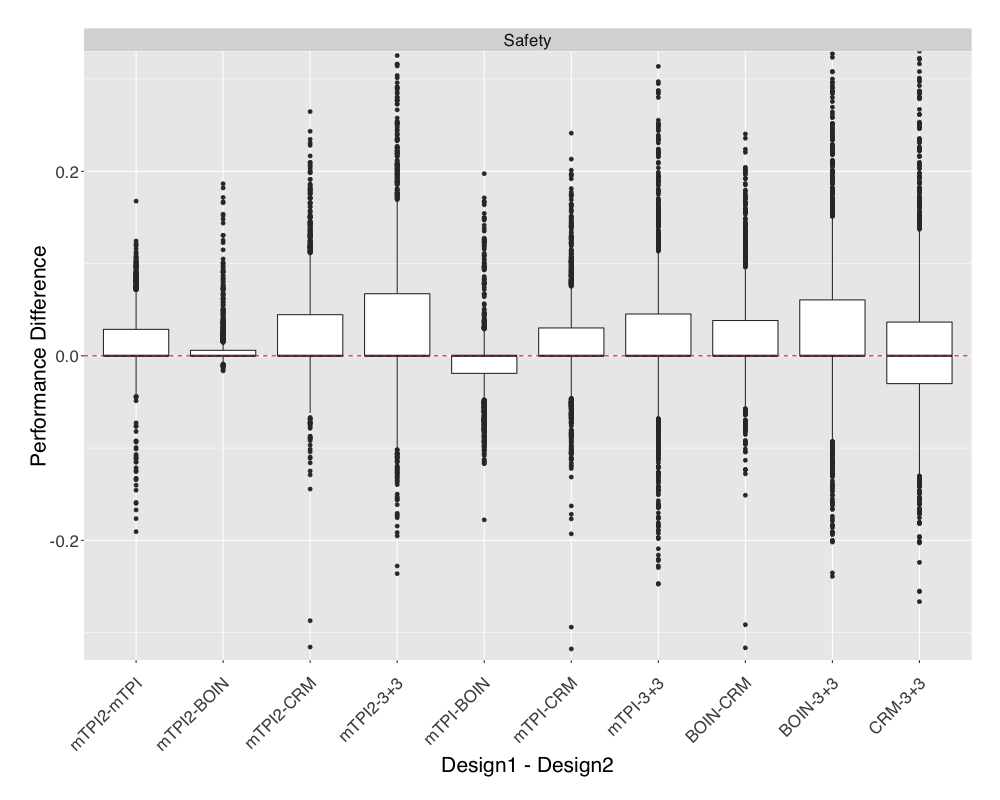} \\
	\includegraphics{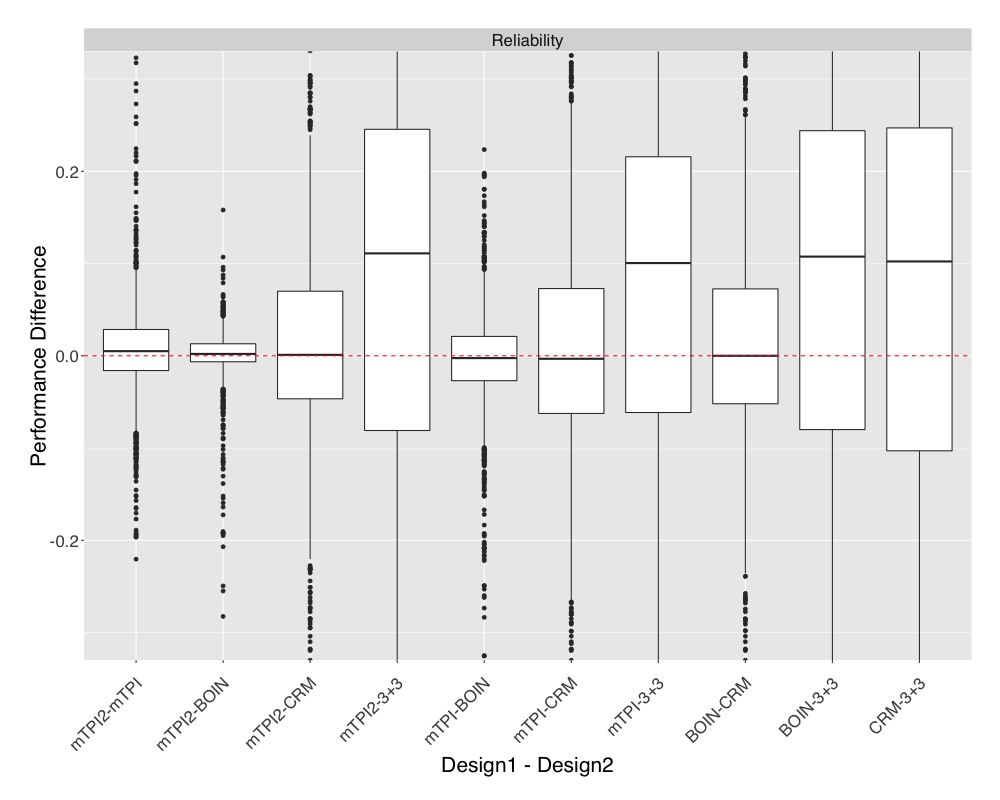}
	\end{tabular}
	}
	\caption{Comparison of {\it safety} and {\it reliability} using the 2,447 crowd-sourcing scenario set one. Five designs are compared which are mTPI2, mTPI, BOIN, CRM, and 3+3. Upper panel [{\it Safety}]: each boxplot describes the differences (Design1 $-$ Design2) in the {\it safety} across all 2,447 scenarios. A value  greater than   zero means Design 1 puts more percentages of patients on doses at or below the true MTD than Design 2. Lower panel [{\it Reliability}]: each boxplot describes the differences (Design1 $-$ Design2) in the {\it reliability} of two designs across all 2,447 scenarios. A value  greater than  zero means Design 1 is more likely to identify the true MTD than Design 2. }\label{fig:res-rs}
	\end{center}
\end{figure}

Figure \ref{fig:res-rs}  presents the comparison results. For ease of
exposition, we only show results from  $\mbox{BOIN}_{\mbox{lambda}}$
so that the  interval boundaries among BOIN, mTPI, and mTPI-2 are the
same. Results including $\mbox{BOIN}_{\mbox{default}}$ and
$\mbox{BOIN}_{\mbox{epsilon}}$ are shown in Appendix A. Overall,
mTPI-2 is the safest design, putting more patients on average at doses
at or below the MTD than the other four methods. BOIN is slightly
worse than the mTPI-2 design  on the safety measure,  followed by mTPI, CRM, and 3+3. Except
for 3+3, all the four model-based designs perform similarly in terms
of {\it reliability}, i.e., the probability of identifying the true
MTD, with mTPI-2 having slight advantage. The 3+3 design is the worst
in the identification of the true MTD. 

\begin{figure}[htbp]
	\begin{center}
	\resizebox{0.9\textwidth}{3.8in}{
	\begin{tabular}{c}
	\includegraphics{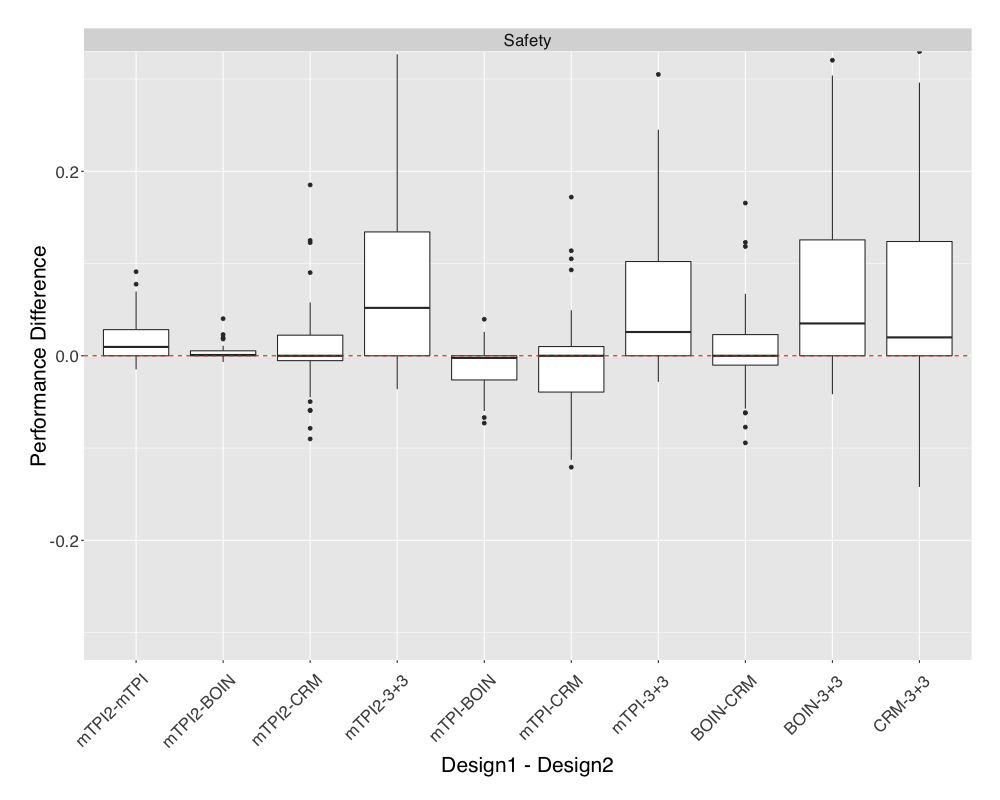} \\
	\includegraphics{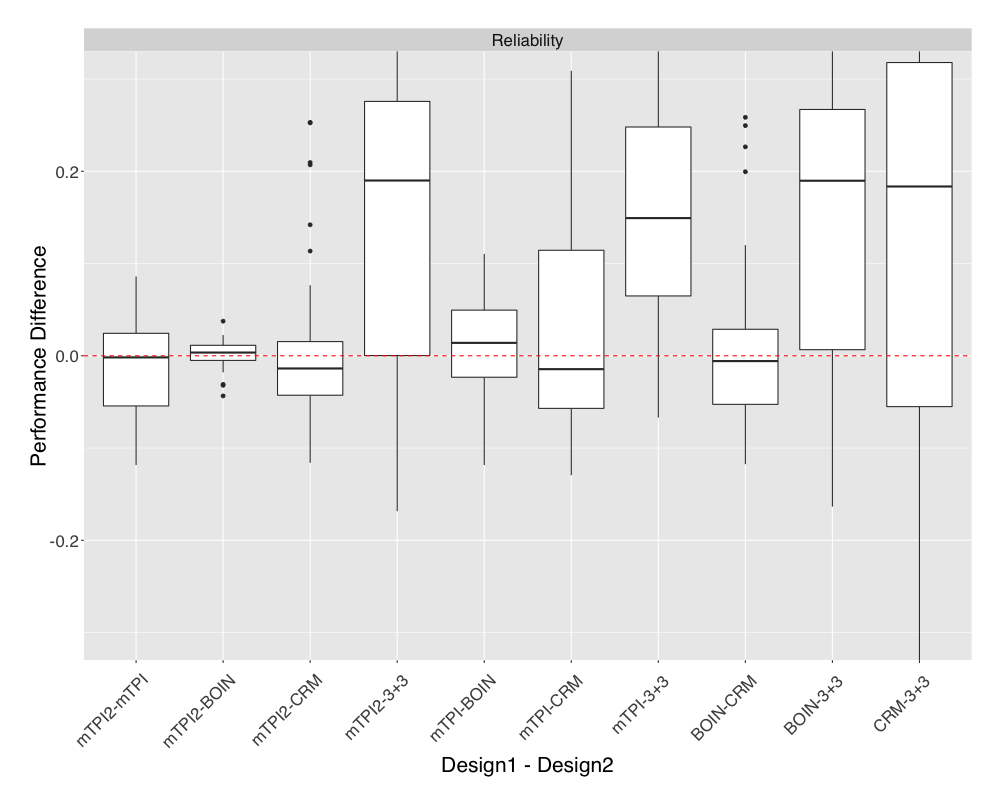}
	\end{tabular}
	}
	\caption{Comparison of {\it safety} and {\it reliability} using the 42 scenarios in scenario set two. Five designs are compared which are mTPI2, mTPI, BOIN, CRM, and 3+3. Upper panel[Safety]: each boxplot describes the differences (Design1 $-$ Design2) in the {\it safety} across all 42 scenarios. A value  greater than  zero means Design 1 puts more percentages of patients on doses at or below the true MTD than Design 2. Lower panel[Reliability]: each boxplot describes the differences (Design1 $-$ Design2) in the {\it reliability} of two designs across all 42 scenarios. A value  greater than  zero means Design 1 is more likely to identify the true MTD than Design 2. }\label{fig:res-jco}
	\end{center}
\end{figure}

\paragraph{Results for set two.} Figure \ref{fig:res-jco} shows the same comparison of the five designs based on the 42 scenarios (Appendix B) in Ji and Wang (2013), with 14 scenarios for each of the three $p_T$ values, 0.1, 0.2, and 0.3. 
As can be seen, mTPI-2 is the safest design while CRM edges mTPI in {\it reliability}, followed closely by mTPI-2 and BOIN.   Again, the 3+3 design is the worst in both safety and reliability.  

\begin{figure}[htbp]
\centering
\begin{tabular}{|c|}\hline\hline
 \includegraphics[scale=0.31]{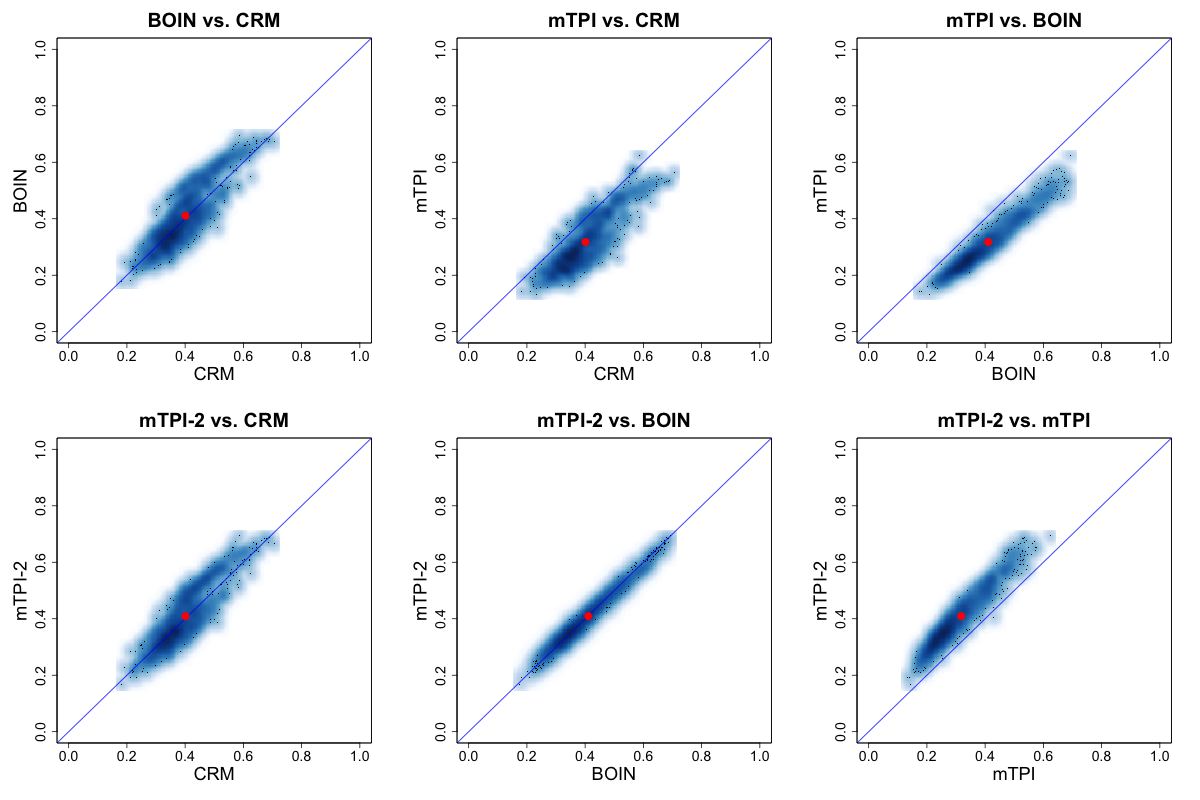} \\ 
 Scatter plots of PCS \\ \hline\hline
	\includegraphics[scale=0.31]{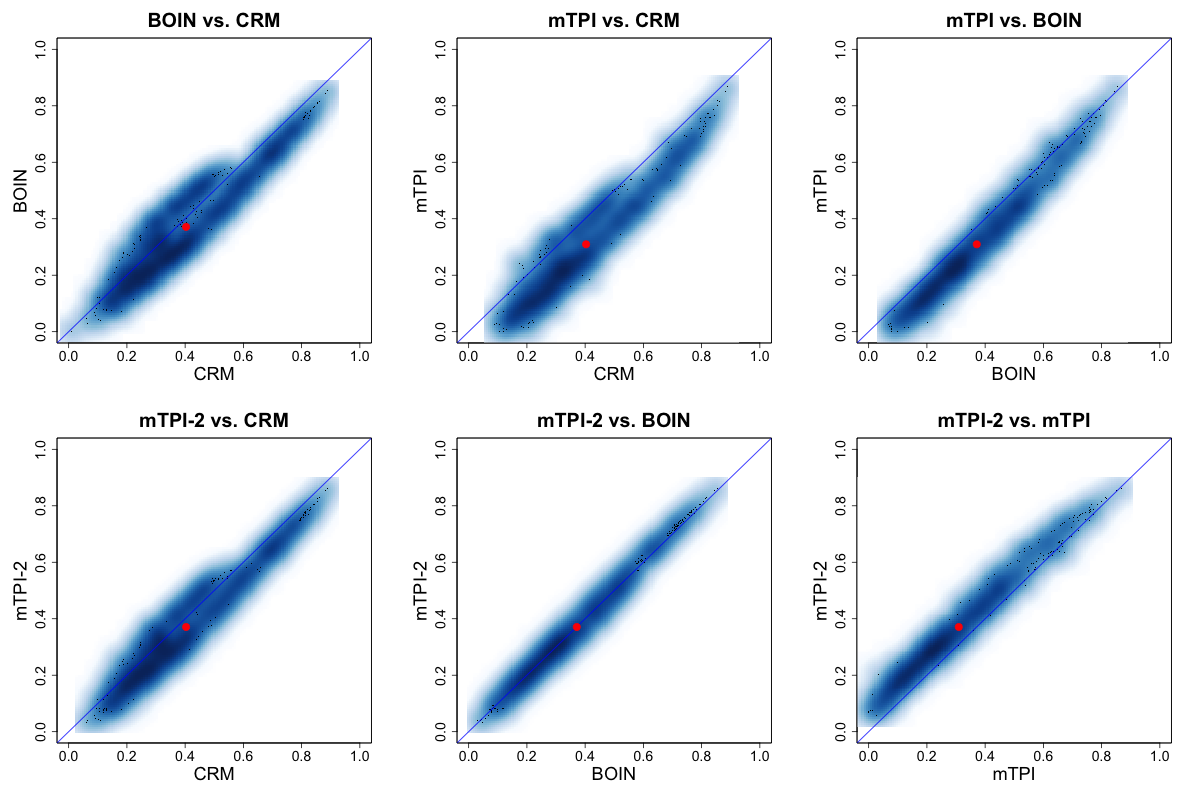} \\ 
	Scatter plots of Accuracy Index for subjects allocation \\ \hline
\end{tabular}
\caption{Smooth scatter plots comparing the mTPI-2, mTPI, CRM and BOIN designs using 1,000 randomly generated scenarios according to Paoletti et al. (2004).   The red dot correspond to the two mean values of the two methods being compared within each plot.  Top panel [PCS]: The percentage of trials of correct selection of the true MTD, which is the same as the reliability criterion. Bottom panel [Accuracy index for subjects allocation]: a criterion $<1$ that measures how accurate patients are allocated surrounding doses near the true MTD. A value closer to 1 means more accurate allocation.}\label{fig:third}
\end{figure}

\paragraph{Results for set three.} 
The purpose of having set three is to examine the performance of all the designs when scenarios are systematically generated from the stochastic model in Paoletti et al. (2004). 
For illustration purpose, we assumed $p_T=0.2$, cohort size 1, and six doses per scenario. We simulated 2,000 trials per scenario and presented the pair-wise comparison of the mTPI-2, mTPI, CRM, and BOIN designs using two criteria 
: 1) PCS: the percentage of trials of correct selection of the true MTD (this is the same as {\it reliability}); 2) Accuracy index for patient allocation   (Cheung, 2011):  a value $<1$ that measures how accurate patients are allocated to the MTD and equals 1 when all the patients are allocated at the true MTD. Results are presented in Figure \ref{fig:third}. Overall, mTPI-2, CRM and BOIN perform very similarly.  As shown in previous work (Horton et al., 2016
) the CRM and BOIN designs perform well in these scenarios. The mTPI design is slightly worse than the other three designs due to its stickiness (Guo et al., 2017) as a result of Occam's razor. 
The mTPI-2 design presents comparable performance. 

\section{Evaluation of Decision Tables}\label{sec:table} 
\begin{figure}[htbp]
  \centering
  \includegraphics[scale=0.35]{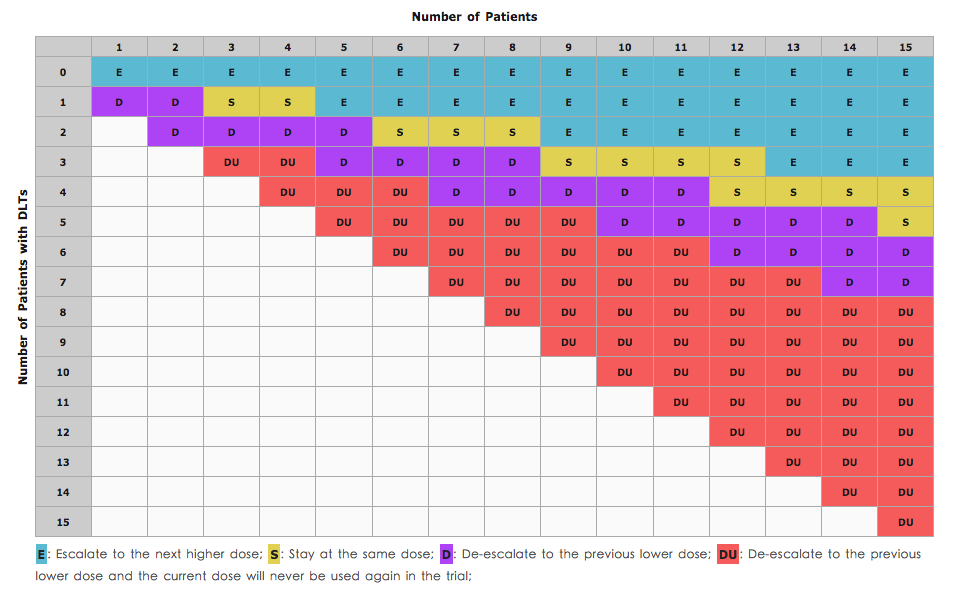}
  \caption{A decision table under the mTPI-2 design for $p_T=0.3$ and $\epsilon_1=\epsilon_2=0.05.$ Each column represents $n$ number of patients treated at the current dose and each row represents $x$ number of patients with DLTs. }\label{fig:mtpi2-table}
\end{figure}

  As in Section \ref{sec:example} and the vast literature,  the
  gold standard of comparing different dose-finding methods is the
  operating characteristics (OC) from running a large number of simulated
  clinical trials on computer, with randomly generated data based on
  prespecified scenarios  that determine the true toxicity
  probabilities of   the  candidate doses. While this provides a
  large-sample (of trials) and population-level assessment on the
  performance of the methods, 
 it does not evaluate designs   in terms of per-trial and per-patient
 behavior. For example, results in Figures \ref{fig:res-rs} compare
 the distributions of the safety and reliability values for each
 design when tested on millions of simulated trials. But for an
 individual investigator, the trial at hand is of the most
 consideration. Practically, these OC results are not well understood
 by the clinical society and are often perceived as a ``black-box''
 operation. Also, due to lack of user-friendly tools, it is
 challenging to generate the OC results for multiple dose-finding
 designs in practice.  One of the key features for the 3+3 design is its
 transparent nature as all the decision rules can be prespecified and
 assessed without the need of running simulation. This same feature is
 also enjoyed by interval-based designs. 
For this reason, we propose to simply evaluate different designs based
on the decision tables that consist of the dose-assignment decisions
of each design. This evaluation can be added to the classical OC
tables. For example, Figure \ref{fig:mtpi2-table} provides a decision table   of mTPI-2  based on $p_T = 0.3$ and $\epsilon_1 = \epsilon_2 = 0.05$ for up to 15 patients. Such a table allows investigators to examine all the possible decisions before the trial starts.  We denote such a table as $\{R_{x,n}(\mbox{mTPI-2}, p_T, \epsilon_1, \epsilon_2)\}$ which consists of decisions of mTPI-2 for values of $p_T$, $\epsilon_{1,2}$, and up to $n$ patients and $x \le n$ possible DLT outcomes. Here $R_{x,n}$ takes three values \{D, S, E\} to denote the three up-and-down dose-assignment decisions. All the interval-based designs, mTPI-2, mTPI, and BOIN can provide such a table prior to the trial starts, and one common feature across the three designs is that the table entry $R_{x,n}$ is fixed given $x$ and $n$ for fixed $p_T$ and $\epsilon_{1,2}$. For example, when 1 out of 3 patients experiences DLT, the table in Figure \ref{fig:mtpi2-table} shows the decision S, which is to stay at the current dose. This decision is applied to any dose level and any trial with $p_T=0.3, \epsilon_1=\epsilon_2=0.5$. 

\begin{figure}[htbp]
\centering
  \includegraphics[scale=0.45]{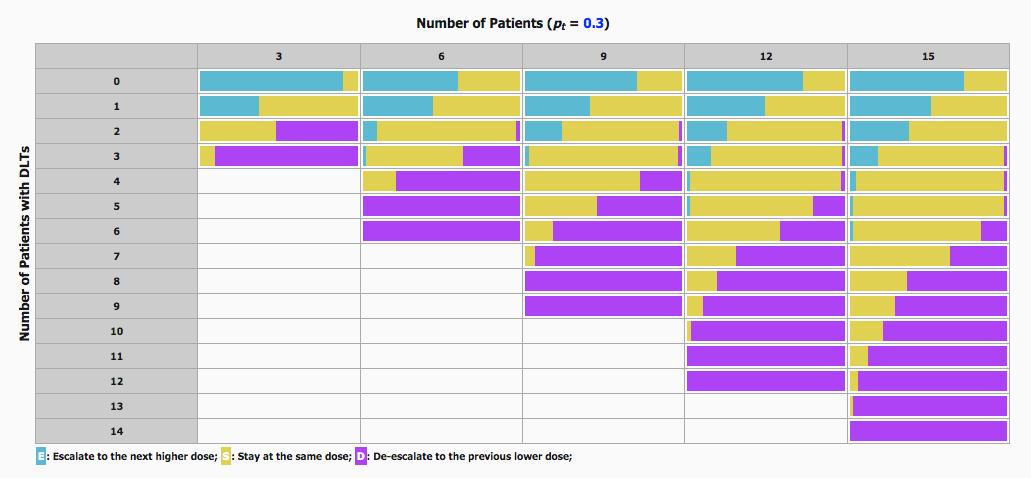}
  \caption{A decision table under the CRM design for $p_T=0.3$. Each
    column represents $n$ number of patients treated at the current
    dose and each row represents $x$ number of patients with DLTs. For
    each entry, the decisions E, S, and D are colored blue, stay, and
    purple, respectively. The length of the colored segments within
    each bar are proportional to the three proportions $(q_{x,n}(E),
    q_{x,n}(S),q_{x,n}(D))$  of the three decisions taken by CRM for a
    given $(x, n)$ data point from a simulation study using the 14 scenarios (Ji and Wang, 2013) and 10,000 simulated trials per scenarios. }\label{fig:crm-table}
\end{figure}

For CRM, the table $\{R_{x,n}(\mbox{CRM}, p_T)\}$ does not depend on $\epsilon_{1,2}$ since CRM does not require input of these two values. More importantly, each entry $R_{x,n}$ is not fixed but a random variable that follows a probability distribution depending on the probability model in the CRM design and the true toxicity probabilities of the doses. For example, consider a trial targeting $p_T=0.3$. If 1 out of 3 patients experiences DLT, CRM will have a large probability (not equal to 1) to S, stay, but also small probabilities to E or D, depending on the patients data on other doses. Unfortunately, the probability distribution of $R_{x,n}$ cannot be derived analytically as it requires integration of all the possible outcomes of the trials that would have 1 DLT out of 3 patients at any dose. Instead, an investigator can obtain a numerical approximation of the random decision table under CRM based on the same   computer   simulated trials for   generating the operating characteristics,   and  examine the   approximated decision  table  carefully before proceeding. To see this, we provide an example next. Suppose we want to design a dose-finding trial with six doses, cohort size 3, and MTD target $p_T=0.3$. We apply CRM to the 14 scenarios in Ji and Wang (2014) with $p_T=0.3$ (Appendix B) and sample size 51. We  simulate 10,000 trials per scenario on computer using the R package {\tt dfCRM} (\url{https://cran.r-project.org/web/packages/dfcrm/index.html}).  We then tabulate the   frequencies of the decisions D, S, or E that CRM takes whenever $x$ out of $n$ patients experience DLT in the simulated trials. These frequencies are the empirical distribution of the random decision $R_{x,n}$ under CRM since they are integrated over the data from  a large number of simulated trials under various scenarios. In other words, this can be considered as a Monte Carlo approximation of the true distribution of $R_{x,n}(\mbox{CRM}, 0.3).$ Note that the actual CRM decisions specify the exact dose level for the next cohort of patients, but these decisions can be easily converted to D, S, or E based on the decided dose level and the current dose level.  Therefore $R_{x,n}$ only takes D, S, or E, and  we obtain three proportions $\{q_{x,n}(D), q_{x,n}(S), q_{x,n}(E)\}$
for each $x$ and $n$, where $q_{x,n}(D) + q_{x,n}(S) + q_{x,n}(E)=1.$ 
 The results are summarized as a decision table in Figure
\ref{fig:crm-table} for up to 15 patients. We removed results for more
than 15 patients for ease of exposition. This table allows investigators to examine the
specific CRM method (in our case, the CRM is the one from the dfCRM R
package) and its performance under the specific trial setting
($p_T=0.3$,   cohort size 3,  and sample size 51) using a specific set
of scenarios. For example, it can be seen from the table that more
than half  of the times the CRM design would stay when 1 out of 3
patients experienced DLT at a given dose (the yellow segment being
longer than the blue segment in the bar for 1 DLT and 3
patients). When 2 out of 3 patients experienced DLT, about half of the
times CRM would de-escalate but the other half of the times it would
stay. \jj These percentages various decisions are important for
investigators and review boards to assess the performance of CRM and
any designs being considered. And importantly, they are more intuitive and transparent
than the OC results. 

\begin{figure}[htbp]
\hspace{-0.2in}
\begin{tabular}{ccc}
  \includegraphics[scale=0.13]{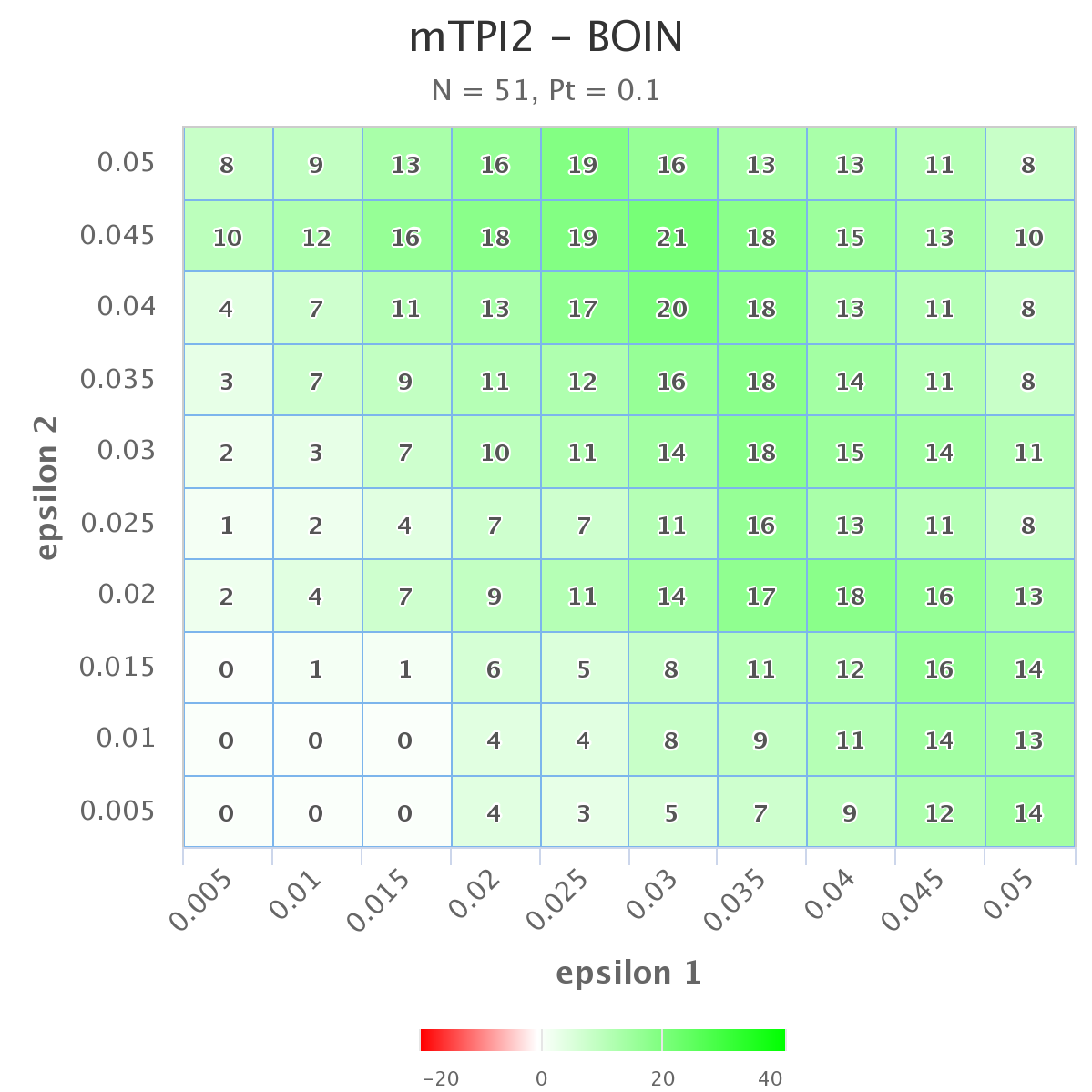} &   \includegraphics[scale=0.13]{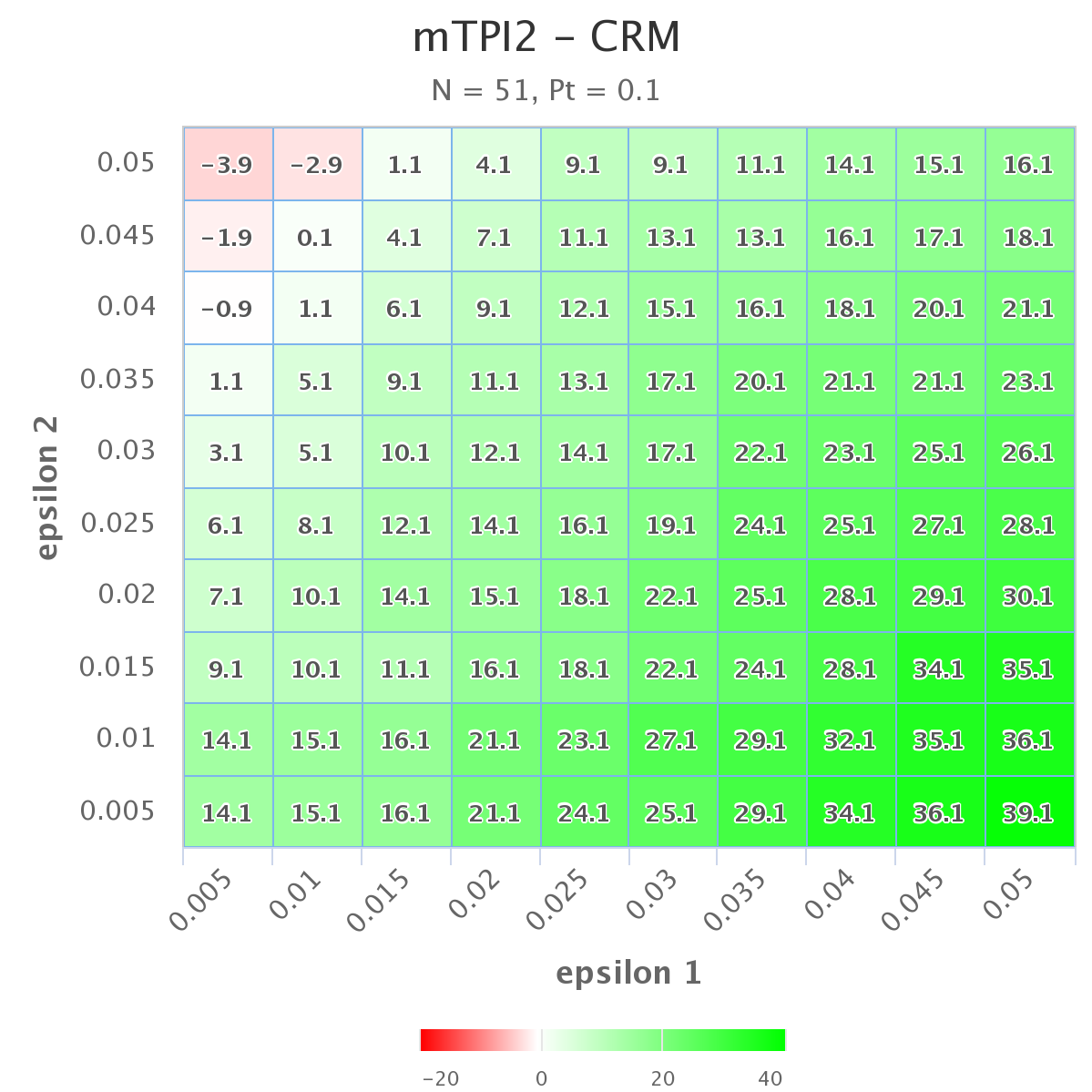} &  \includegraphics[scale=0.13]{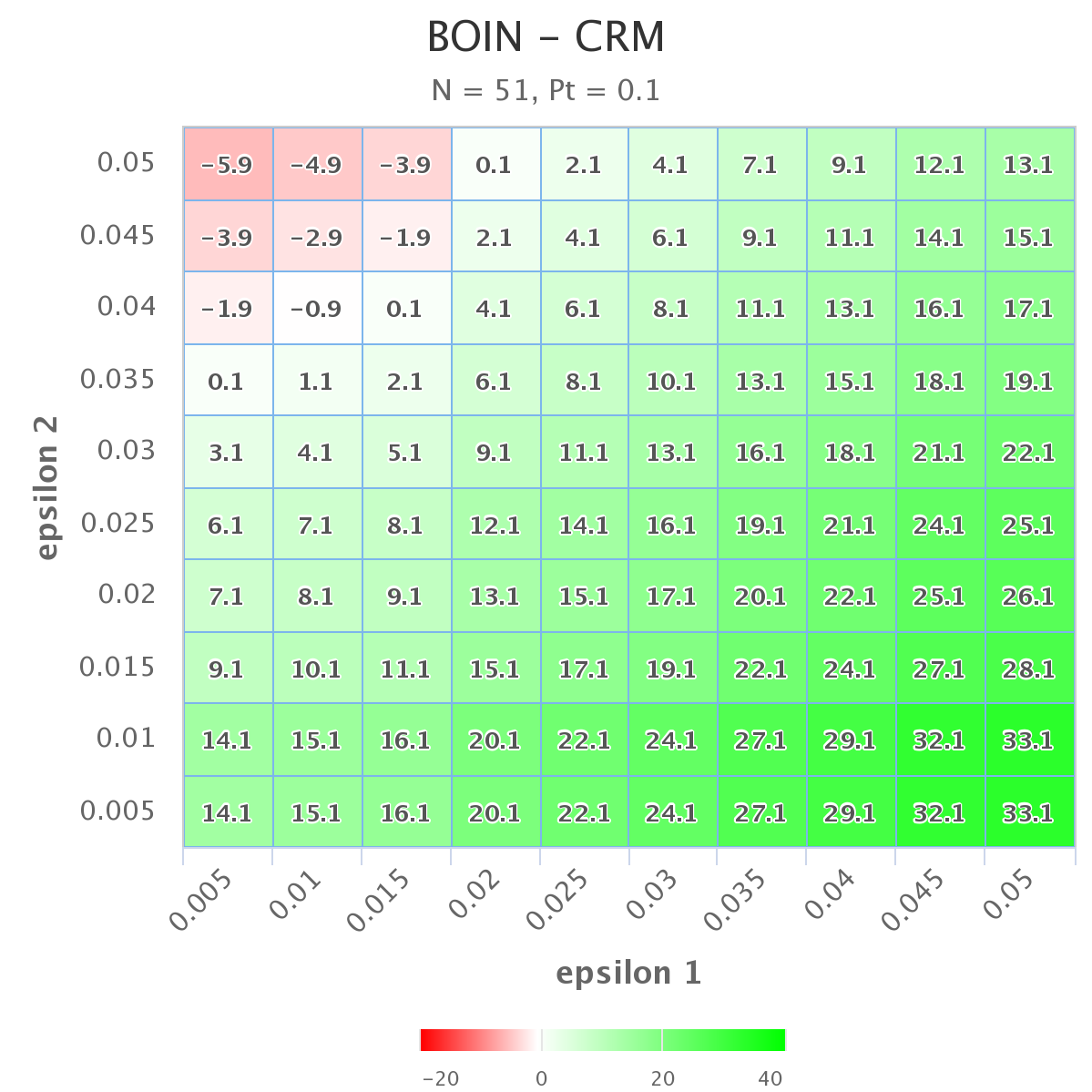} \\
  \includegraphics[scale=0.13]{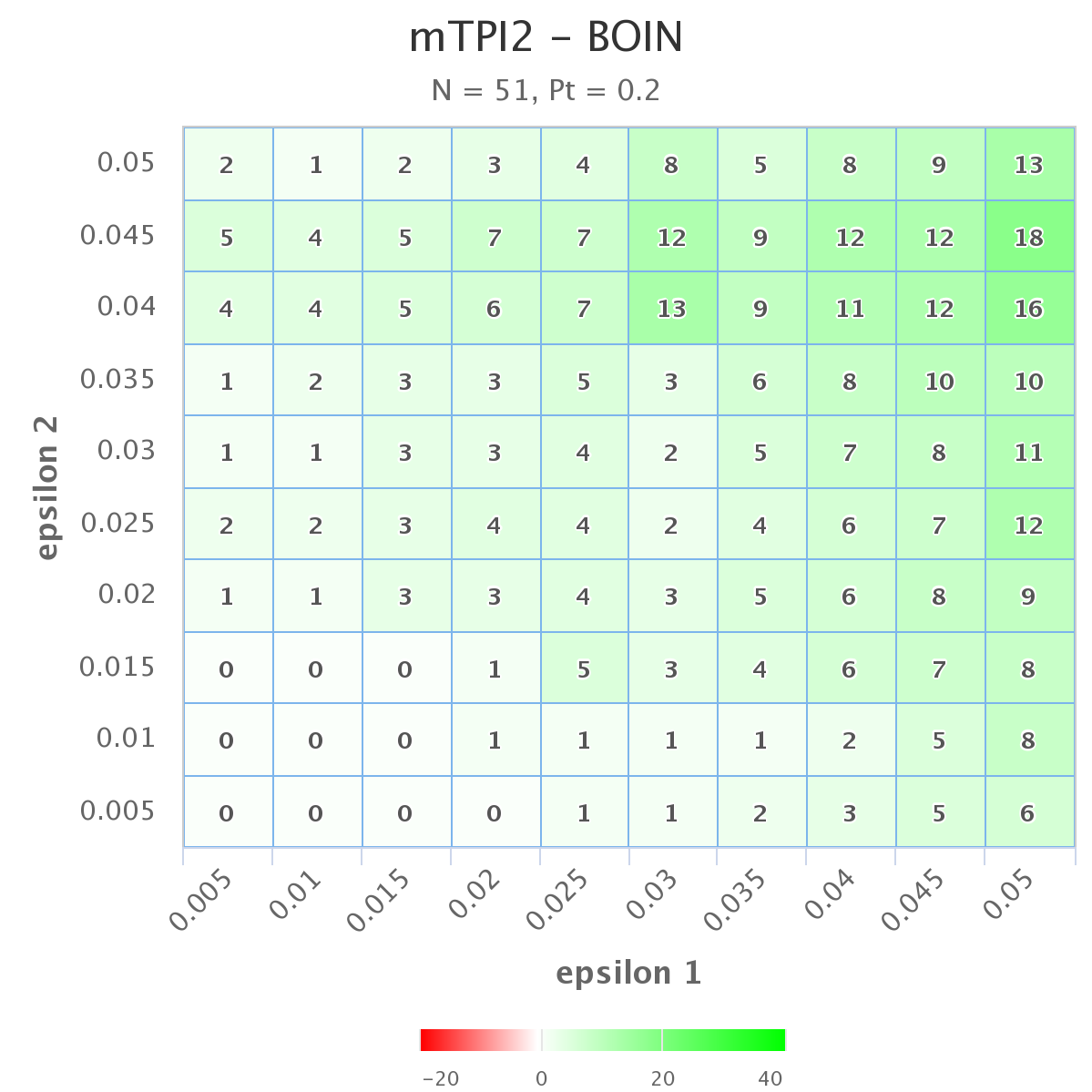} &   \includegraphics[scale=0.13]{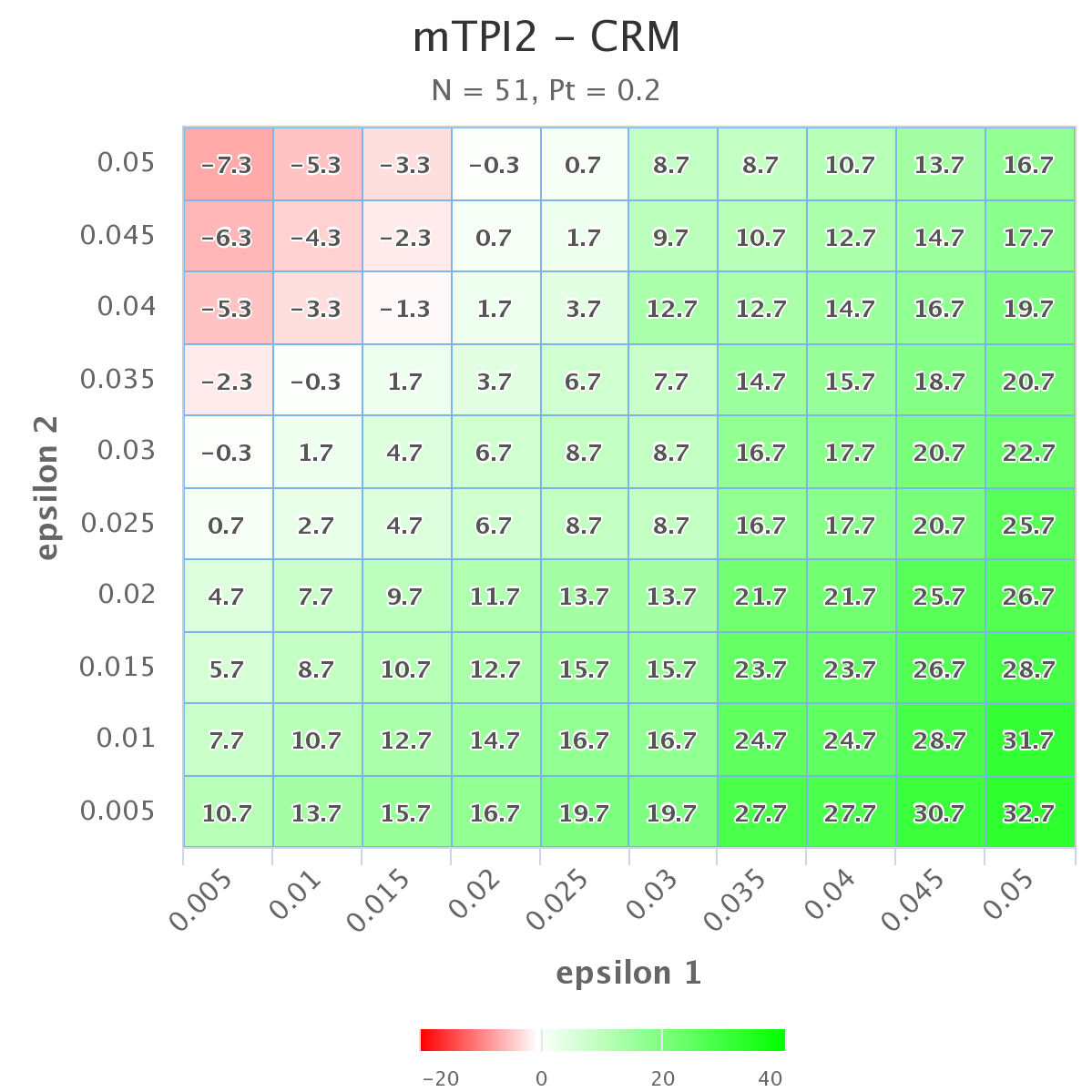} &  \includegraphics[scale=0.13]{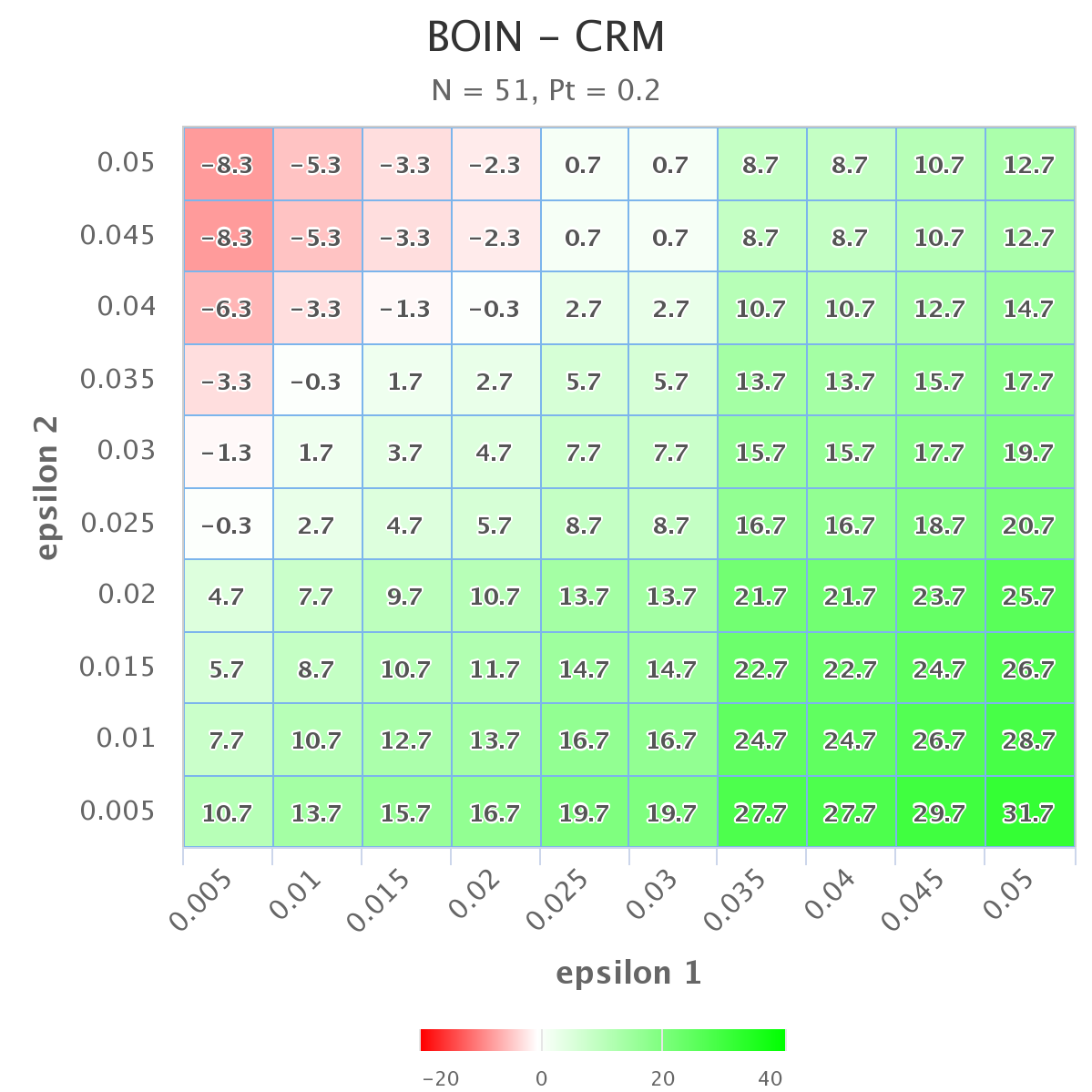} \\
  \includegraphics[scale=0.13]{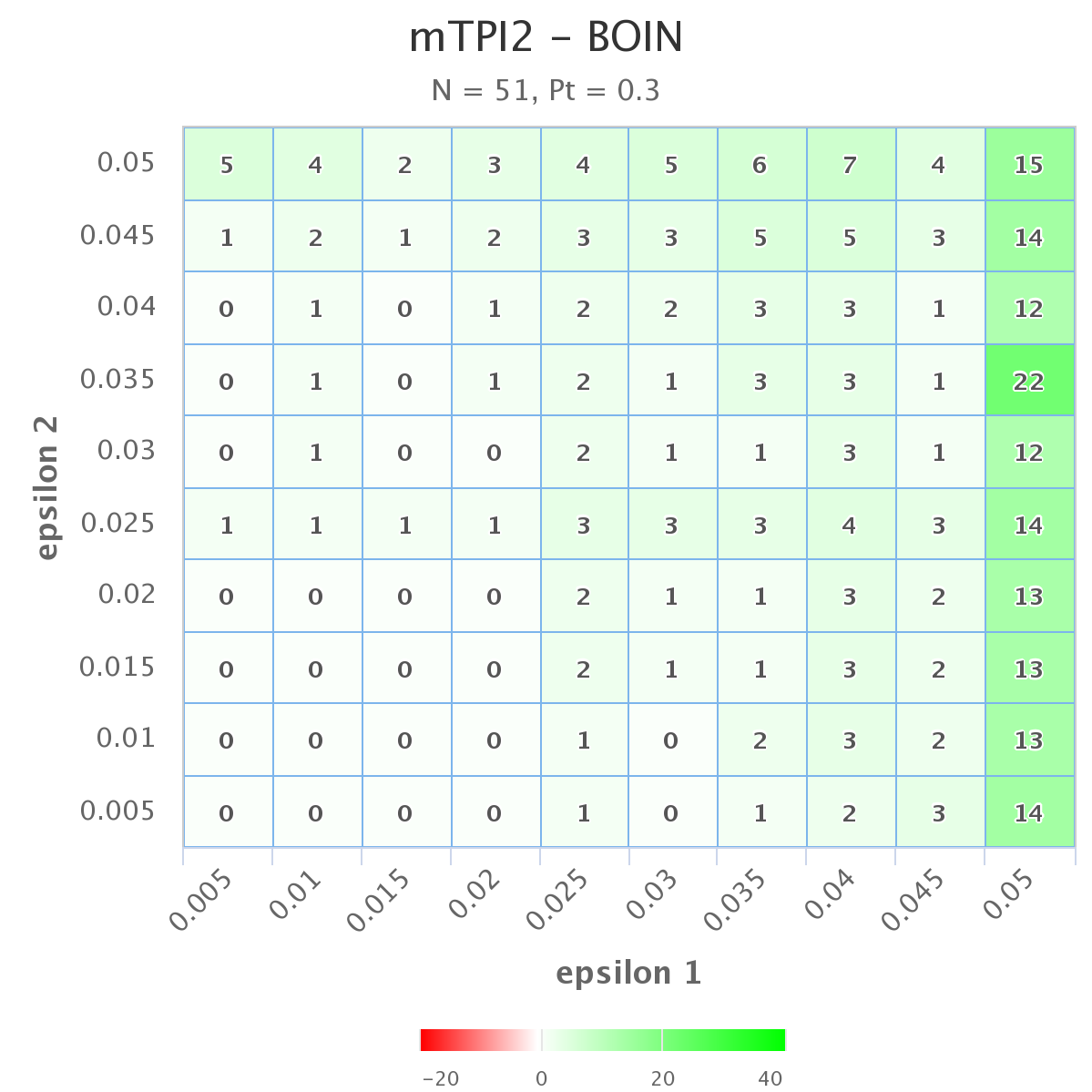} &   \includegraphics[scale=0.13]{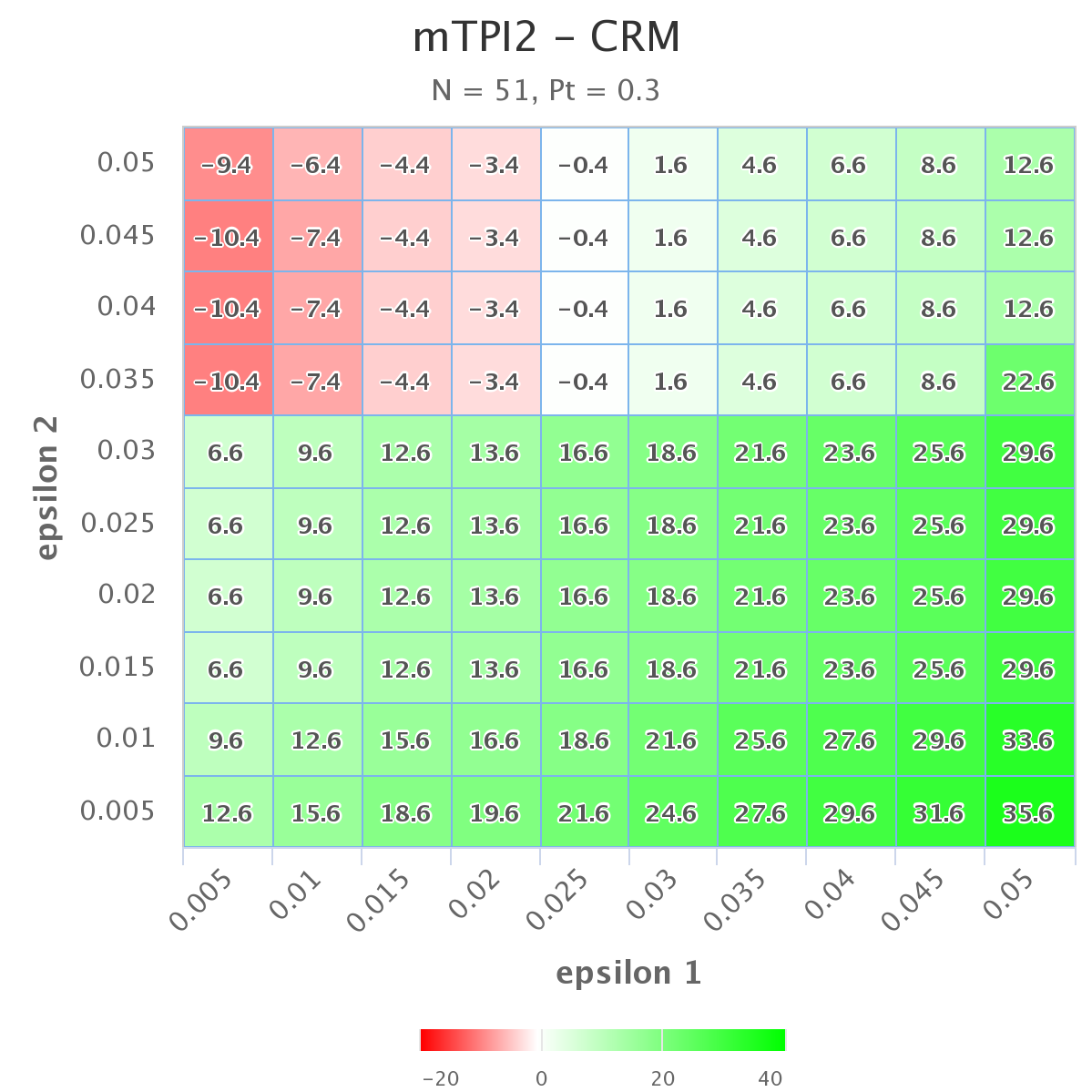} &  \includegraphics[scale=0.13]{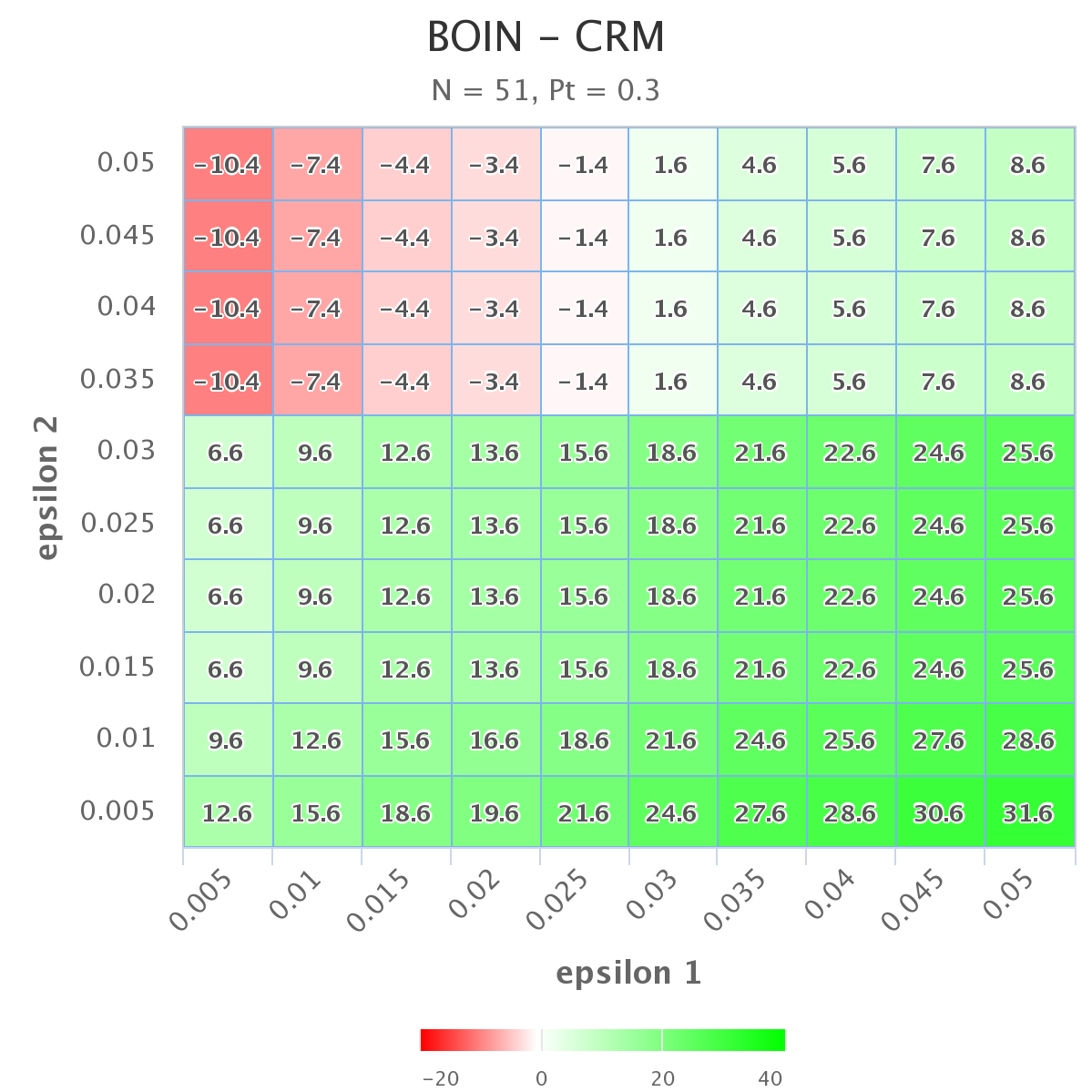} \\
  \end{tabular}
  \caption{Pairwise difference decision tables of the mTPI-2, CRM and
    BOIN$_{\mbox{lambda}}$ designs for different $p_T$ values of 0.1,
    0.2, and 0.3, and different $\epsilon_{1,2}$ values. The decisions
    E, S, and D are coded as 1, 2, and 3, respectively. Each table
    entry is the sum of the differences (Design1 - Design2) of the
    decisions $R_{x,n}(Design1,p_T,\epsilon_1, \epsilon_2) -
    R_{x,n}(Design2,p_T,\epsilon_1, \epsilon_2)$ between two designs
    for $n=51$ patients. For CRM, we use $E(R_{x,n}(CRM, p_T))$ in the
    calculation.   A positive/negative value or green/red color means Design1 is more likely to
    de-escalate/escalate.  }\label{fig:decision-diff}
\end{figure}

As a comparison, we also obtain the fixed decision tables (not shown)
for a sample size of $n=51$ patients for the mTPI-2 and
BOIN$_{\mbox{lambda}}$ designs given $p_T=0.3$ value and a pair of
$\epsilon_{1,2}$ values, denoted as $\{R_{x,n}(\mbox{mTPI2}, p_T,
\epsilon_1, \epsilon_2)\}$ and $\{R_{x,n}(\mbox{BOIN}, p_T,
\epsilon_1, \epsilon_2)\}$. Since these tables not only depend on
$p_T$ values, but also $\epsilon_{1,2}$, we vary both $\epsilon$'s
from 0.005 to 0.05. For CRM, for each $(x, n)$ value we obtain three
proportions $\{q_{x,n}(D), q_{x,n}(S), q_{x,n}(E)\}$ which represent
the proportions of the three decisions. As a summary, we compute the mean decision score
defined as $E(R_{x,n}, p_T)=1*q_{x,n}(E) + 2*q_{x,n}(S) +
3*q_{x,n}(D)$ for each $x$ and $n$ value, where values $1$, $2$,
and $3$ are assigned to decision $E$, $S$, and $D$,
respectively. Therefore, a higher value is associated with a more
conservative decision. Figure \ref{fig:decision-diff} shows the sum of  the  differences (Design1 - Design2)  of $R$ for mTPI-2 and BOIN and $E(R)$ for CRM between any two designs for each of  three $p_T$ values, $0.1, 0.2,$ and $0.3$. For example,  the top left panel presents the differences  $\sum_{n=1}^{51}\sum_{x=1}^n \{R_{x,n}(\mbox{mTPI-2}, 0.1,, \epsilon_1, \epsilon_2) - R_{x,n}(\mbox{BOIN}, 0.1, \epsilon_1, \epsilon_2)\}$ for all the $\epsilon_{1,2}$ values.  A positive or negative value in the plot implies that Design1 is more likely to de-escalate or escalate, respectively.    Overall mTPI-2 appear to  have more de-escalation decisions than the other two designs in most cases,    and the two interval-based designs do not differ much in general. The differences of the two designs get larger for larger $\epsilon$ values.  CRM  is in general more aggressive in its decisions than mTPI-2 and BOIN. However, when $\epsilon_2$ is large and $\epsilon_1$ is small, the equivalence interval $(p_T - \epsilon_1, p_T + \epsilon_2)$ becomes more skewed towards right, which means mTPI-2 and BOIN will be less likely to de-escalate and more likely to escalate. Therefore, we see the decisions of mTPI-2 and BOIN are more aggressive than CRM in these $\epsilon$ values (upper left corner of the heatmaps  in Figure \ref{fig:decision-diff}).

\section{Discussion}
 
We
show using 
 crowd-sourcing results that mTPI, mTPI-2, and
BOIN designs   are superior than 3+3 and  compare well with other non-interval designs, such as CRM. Due   to  the simplicity in the implementation of the interval
designs, we recommend the use of these designs for practical
trials. In particular, mTPI-2 seems to stand out with its rigorous
theoretical development, superior numeric performance,  simplicity
in the implementation, and  adherence to the ethical principle   in its dose escalation decisions. 

The interval designs can be categorized as the iDesigns, including TPI, mTPI, and mTPI-2, and IB-Designs including CCD and BOIN. They differ in the statistical inference and underlying decision strategies. Interestingly, they are very similar in implementation which produces a fixed decision table and is attractive to practitioners due to its simplicity and transparency. The use of deterministic decisions allow their decision tables to be calculated and  examined  prior to the trial onset, allowing investigators to see the behavior of the designs given various patient outcomes in terms of $x$ number of DLTs out of $n$ patients treated at a dose.   The mTPI-2 design appears to have the best overall performance in our comparison using the the 42 scenarios in Ji and Wang (2013). The performance of mTPI-2 is similar to BOIN on overall reliability or PCS using the randomly generated scenarios. However, the decision tables give clear contrast that mTPI-2 is  a safer  design.  

We also show that running a large number of computer simulated clinical trials and comparing operating characteristics of the designs is not sufficient to fully assess dose-finding designs in practice. We recommend to examine the decision tables, even for methods like CRM when its decision tables are not deterministic. Since a large number of simulated trials are usually conducted on computer as a standard practice, one can quickly generate the empirical distributions of the CRM decisions $R_{x,n}$ as in Figure \ref{fig:crm-table} to assess the performance of CRM for a given pair of $(x, n)$ values. Decision tables and operating characteristics tables will jointly allow investigators to evaluate the population-level (across many simulated trials) and individual-level (across patients) behavior of a design. Therefore, both of them should be reviewed  when these designs are considered for a  practical trial. 

\section*{Acknowledgement}
This is a collaborative research with Dr. Sue-Jane Wang from US FDA. 
She provided numerous insightful critiques, did some technical
investigations and gave constructive inputs throughout the
research. We sincerely thank her for her professional expertise and
scientific collaboration. 


\bibliographystyle{apa}
\bibliography{references2}

\begin{thebibliography}{}

\bibitem[\protect\astroncite{Akaike}{1974}]{akaike1974new}
Akaike, H. (1974).
\newblock {A new look at the statistical model identification}.
\newblock {\em IEEE transactions on automatic control}, 19(6):716--723.

\bibitem[\protect\astroncite{Babb et~al.}{1998}]{babb1998cancer}
Babb, J., Rogatko, A., and Zacks, S. (1998).
\newblock {Cancer phase I clinical trials: efficient dose escalation with
  overdose control}.
\newblock {\em Statistics in medicine}, 17(10):1103--1120.

\bibitem[\protect\astroncite{Berger}{2013}]{berger2013statistical}
Berger, J.~O. (2013).
\newblock {\em Statistical decision theory and Bayesian analysis}.
\newblock Springer Science \& Business Media.

\bibitem[\protect\astroncite{Blanchard and
  Longmate}{2011}]{blanchard2011toxicity}
Blanchard, M.~S. and Longmate, J.~A. (2011).
\newblock Toxicity equivalence range design (teqr): a practical phase i design.
\newblock {\em Contemporary clinical trials}, 32(1):114--121.

\bibitem[\protect\astroncite{Cheung}{2011}]{cheung2011dose}
Cheung, Y.~K. (2011).
\newblock {\em {Dose finding by the continual reassessment method}}.
\newblock CRC Press.

\bibitem[\protect\astroncite{Cheung and Chappell}{2002}]{cheung:2002}
Cheung, Y.~K. and Chappell, R.~A. (2002).
\newblock A simple technique to evaluate model sensitivity in the continual
  reassessment method.
\newblock {\em Biometrics}, 58(3):671--674.

\bibitem[\protect\astroncite{Clertant and O'Quigley}{2017}]{Clertant:2017}
Clertant, M. and O'Quigley, J. (2017).
\newblock Semiparametric dose finding methods.
\newblock {\em Journal of the Royal Statistical Society, Series B},
  79(3):10.1111/rssb.12229.

\bibitem[\protect\astroncite{Gezmu and Flournoy}{2006}]{gezmu2006group}
Gezmu, M. and Flournoy, N. (2006).
\newblock Group up-and-down designs for dose-finding.
\newblock {\em Journal of statistical planning and inference},
  136(6):1749--1764.

\bibitem[\protect\astroncite{Goodman et~al.}{1995}]{goodman1995some}
Goodman, S.~N., Zahurak, M.~L., and Piantadosi, S. (1995).
\newblock {Some practical improvements in the continual reassessment method for
  phase I studies}.
\newblock {\em Statistics in medicine}, 14(11):1149--1161.

\bibitem[\protect\astroncite{Guo et~al.}{2017}]{guo2016bayesian}
Guo, W., Wang, S.-J., Yang, S., Lin, S., and Ji, Y. (2017).
\newblock {A Bayesian Interval Dose-Finding Design Addressing Ockham's Razor:
  mTPI-2}.
\newblock {\em Contemprary Clinical Trials}, in press.

\bibitem[\protect\astroncite{Ivanova et~al.}{2007}]{ivanova2007cumulative}
Ivanova, A., Flournoy, N., and Chung, Y. (2007).
\newblock {Cumulative cohort design for dose-finding}.
\newblock {\em Journal of Statistical Planning and Inference},
  137(7):2316--2327.

\bibitem[\protect\astroncite{Jefferys and Berger}{1992}]{jefferys1992ockham}
Jefferys, W.~H. and Berger, J.~O. (1992).
\newblock Ockham's razor and bayesian analysis.
\newblock {\em American Scientist}, 80(1):64--72.

\bibitem[\protect\astroncite{Ji et~al.}{2007a}]{ji2007dose}
Ji, Y., Li, Y., and Bekele, B.~N. (2007a).
\newblock {Dose-finding in phase I clinical trials based on toxicity
  probability intervals}.
\newblock {\em Clinical Trials}, 4(3):235--244.

\bibitem[\protect\astroncite{Ji et~al.}{2007b}]{ji2007bayesian}
Ji, Y., Li, Y., and Yin, G. (2007b).
\newblock {Bayesian dose finding in phase I clinical trials based on a new
  statistical framework}.
\newblock {\em Statistica Sinica}, pages 531--547.

\bibitem[\protect\astroncite{Ji et~al.}{2010}]{ji2010modified}
Ji, Y., Liu, P., Li, Y., and Bekele, B.~N. (2010).
\newblock {A modified toxicity probability interval method for dose-finding
  trials}.
\newblock {\em Clinical Trials}, page 1740774510382799.

\bibitem[\protect\astroncite{Ji and Wang}{2013}]{ji2013modified}
Ji, Y. and Wang, S.-J. (2013).
\newblock {Modified toxicity probability interval design: a safer and more
  reliable method than the 3+ 3 design for practical phase I trials}.
\newblock {\em Journal of Clinical Oncology}, 31(14):1785--1791.

\bibitem[\protect\astroncite{Liu and Yuan}{2015}]{liu2015bayesian}
Liu, S. and Yuan, Y. (2015).
\newblock {Bayesian optimal interval designs for phase I clinical trials}.
\newblock {\em Journal of the Royal Statistical Society: Series C (Applied
  Statistics)}, 64(3):507--523.

\bibitem[\protect\astroncite{O'Quigley et~al.}{1990}]{o1990continual}
O'Quigley, J., Pepe, M., and Fisher, L. (1990).
\newblock {Continual reassessment method: a practical design for phase 1
  clinical trials in cancer}.
\newblock {\em Biometrics}, pages 33--48.

\bibitem[\protect\astroncite{Paoletti et~al.}{2004}]{paoletti2004design}
Paoletti, X., O'Quigley, J., and Maccario, J. (2004).
\newblock {Design efficiency in dose finding studies}.
\newblock {\em Computational statistics \& data analysis}, 45(2):197--214.

\bibitem[\protect\astroncite{Schwarz et~al.}{1978}]{schwarz1978estimating}
Schwarz, G. et~al. (1978).
\newblock {Estimating the dimension of a model}.
\newblock {\em The annals of statistics}, 6(2):461--464.

\bibitem[\protect\astroncite{Storer}{1989}]{storer1989design}
Storer, B.~E. (1989).
\newblock {Design and analysis of phase I clinical trials}.
\newblock {\em Biometrics}, pages 925--937.

\bibitem[\protect\astroncite{Stylianou and Flournoy}{2002}]{stylianou2002dose}
Stylianou, M. and Flournoy, N. (2002).
\newblock {Dose finding using the biased coin Up-and-Down design and isotonic
  regression}.
\newblock {\em Biometrics}, 58(1):171--177.

\bibitem[\protect\astroncite{Yang et~al.}{2015}]{yang2015integrated}
Yang, S., Wang, S.-J., and Ji, Y. (2015).
\newblock {An integrated dose-finding tool for phase I trials in oncology}.
\newblock {\em Contemporary clinical trials}, 45:426--434.

\bibitem[\protect\astroncite{Yuan et~al.}{2016}]{yuan2016bayesian}
Yuan, Y., Hess, K.~R., Hilsenbeck, S.~G., and Gilbert, M.~R. (2016).
\newblock {Bayesian optimal interval design: a simple and well-performing
  design for phase I oncology trials}.
\newblock {\em American Association for Cancer Research}, 22(17):4291--4301.

\end{thebibliography}

\clearpage
\newpage

\section*{Appendix A}
We present the pair-wise comparison of mTPI-2, mTPI, BOIN, CRM, and 3+3, with BOIN having three versions, $\mbox{BOIN}_{\mbox{default}},$ $\mbox{BOIN}_{\mbox{epsilon}},$ $\mbox{BOIN}_{\mbox{lambda}}.$ The $\mbox{BOIN}_{\mbox{default}}$ and $\mbox{BOIN}_{\mbox{lambda}}$ versions are similar. 

\begin{figure}[htbp]
	\begin{center}
	\resizebox{0.9\textwidth}{3.5in}{
	\begin{tabular}{c}
	\includegraphics{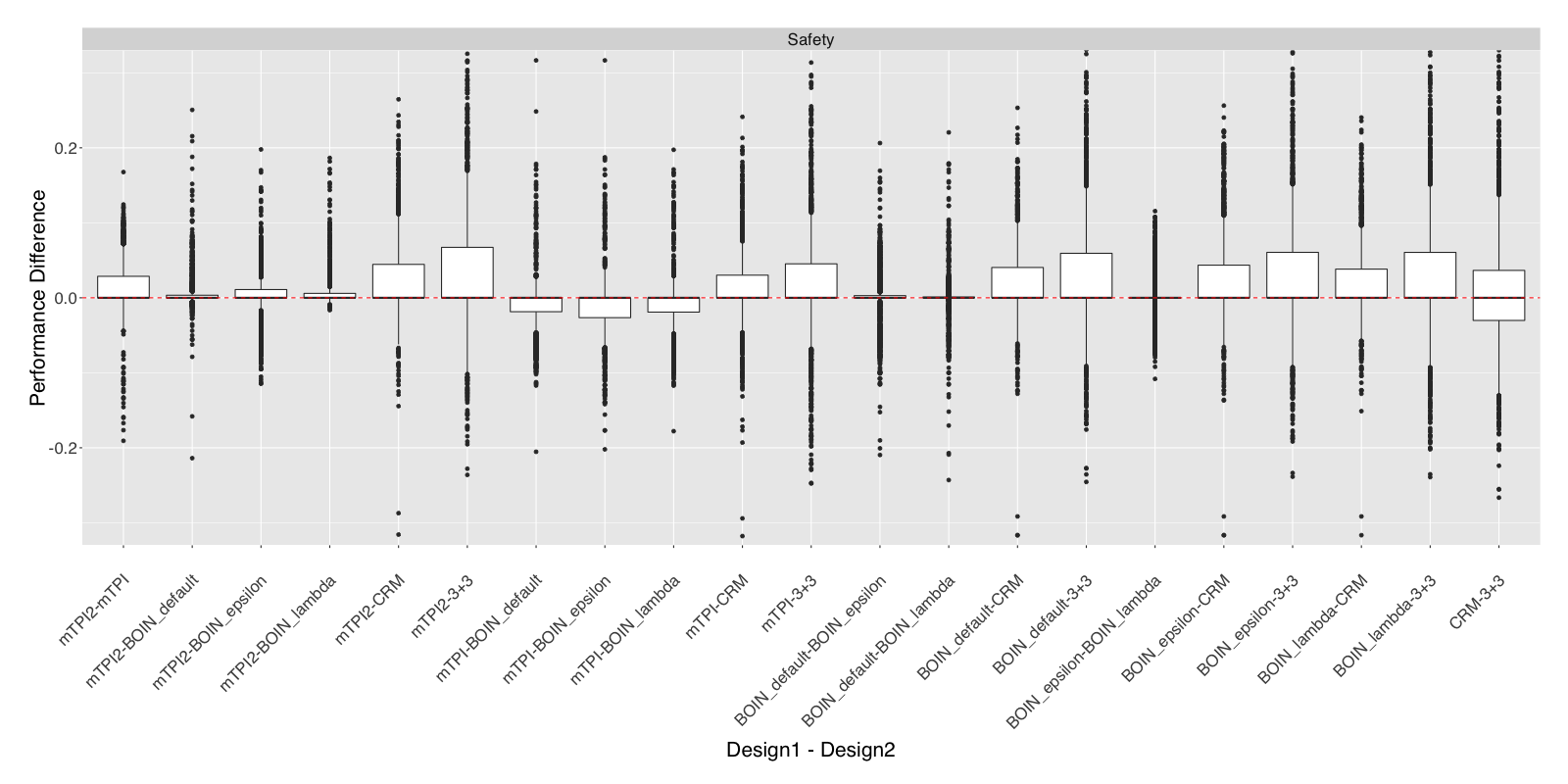} \\
	\includegraphics{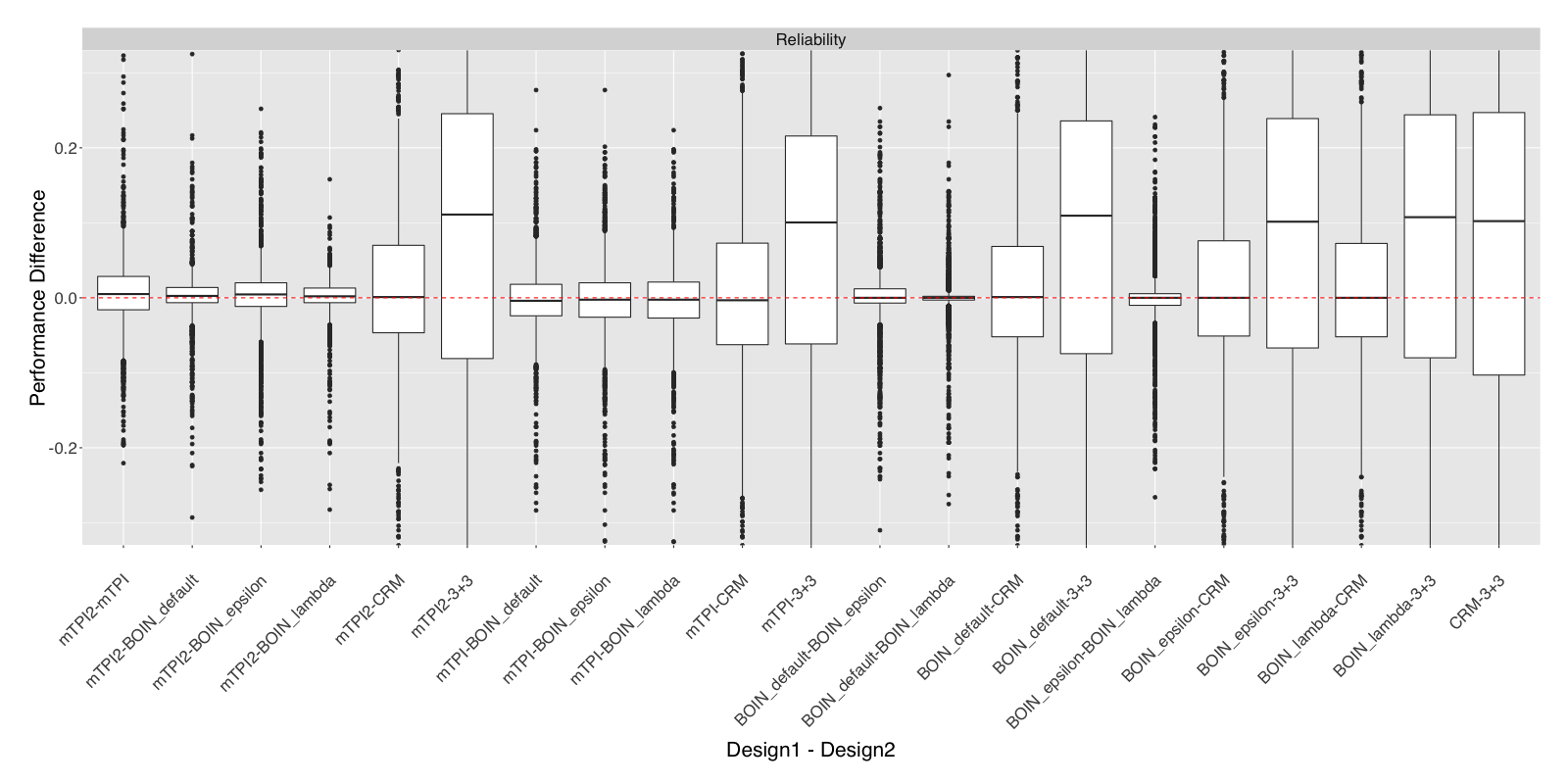}
	\end{tabular}
	}
	\caption{Comparison of {\it safety} using the 2,447 crowd-sourcing scenario set one. Five designs are compared which are mTPI-2, mTPI, BOIN, CRM, and 3+3, with BOIN having three versions $\mbox{BOIN}_{\mbox{default}},$ $\mbox{BOIN}_{\mbox{epsilon}},$ $\mbox{BOIN}_{\mbox{lambda}}.$ Upper panel[Safety]: each boxplot describes the differences (Design1 $-$ Design2) in the {\it safety} across all 2,447 scenarios. A value  greater than   zero means Design 1 puts more percentages of patients on doses at or below the true MTD than Design 2. Lower panel[Reliability]: each boxplot describes the differences (Design1 $-$ Design2) in the {\it reliability} of two designs across all 2,447 scenarios. A value  greater than  zero means Design 1 is more likely to identify the true MTD than Design 2. }\label{fig:res-rs-all}
	\end{center}
\end{figure}

\clearpage
\newpage

\section*{Appendix B}
\singlespacing
The 42 scenarios in Ji and Wang (2013) for $p_T=0.1, 0.2,$ and $0.3$.
\begin{table}[htbp]
{\footnotesize
\begin{center}
\resizebox{0.545\textwidth}{!}{
\begin{tabular}{|c|cccccc|} \hline
Scenario \# & Dose 1 & Dose 2 & Dose 3 & Dose 4 & Dose 5 & Dose 6 \\ \hline
\multicolumn{7}{l}{} \\
\multicolumn{7}{l}{{\bf $p_T$ = 0.1}} \\ \hline
1& 0.04	&	0.05	&	0.06	&	0.07	&	0.08
&	0.09 \\
2& 0.15	&	0.2	&	0.25	&	0.3	&	0.35	&	0.4 \\
3& 0.01	&	0.1	&	0.2	&	0.25	&	0.3	&	0.35 \\
4& 0.01	&	0.02	&	0.03	&	0.04	&	0.1	&	0.25 \\
5& 0.05	&	0.4	&	0.5	&	0.6	&	0.65	&	0.7 \\
6& 0.01	&	0.03	&	0.05	&	0.4	&	0.5	&	0.6 \\
7& 0.01	&	0.02	&	0.03	&	0.04	&	0.05	&	0.4 \\
8& 0.09	&	0.11	&	0.13	&	0.15	&	0.17	&	0.19 \\
9& 0.05	&	0.07	&	0.09	&	0.11	&	0.13	&	0.15 \\
10& 0.01	&	0.03	&	0.05	&	0.07	&	0.09	&	0.11 \\
11& 0.02	&	0.04	&	0.08	&	0.12	&	0.17	&	0.25 \\
12& 0.02	&	0.04	&	0.07	&	0.1	&	0.15	&	0.2 \\
13& 0.1	&	0.15	&	0.2	&	0.25	&	0.3	&	0.35 \\
14& 0.01	&	0.03	&	0.05	&	0.06	&
0.08	&	0.1 \\ \hline
\multicolumn{7}{l}{} \\
\multicolumn{7}{l}{{\bf $p_T$ = 0.2}} \\ \hline
1& 0.02	&	0.05	&	0.08	&	0.11	&	0.14	&	0.17 \\
2& 0.25	&	0.35	&	0.4	&	0.5	&	0.6	&	0.7 \\
3& 0.01	&	0.2	&	0.4	&	0.6	&	0.8	&	0.95 \\
4& 0.04	&	0.06	&	0.08	&	0.1	&	0.2	&	0.5 \\
5& 0.05	&	0.5	&	0.8	&	0.9	&	0.95	&	0.99 \\
6& 0.01	&	0.05	&	0.1	&	0.5	&	0.7	&	0.9 \\
7& 0.01	&	0.03	&	0.07	&	0.1	&	0.15	&	0.7 \\
8& 0.19	&	0.21	&	0.23	&	0.25	&	0.27	&	0.29 \\
9& 0.15	&	0.17	&	0.19	&	0.21	&	0.23	&	0.25 \\
10& 0.11	&	0.13	&	0.15	&	0.17	&	0.19	&	0.21 \\
11& 0.05	&	0.11	&	0.17	&	0.23	&	0.29	&	0.35 \\
12& 0.05	&	0.1	&	0.15	&	0.2	&	0.3	&	0.4 \\
13& 0.2	&	0.25	&	0.3	&	0.35	&	0.4	&	0.45 \\
14& 0.05	&	0.08	&	0.11	&	0.14	&
0.17	&	0.2 \\ \hline
\multicolumn{7}{l}{} \\
\multicolumn{7}{l}{{\bf $p_T$ = 0.3}} \\ \hline
1& 0.02	&	0.05	&	0.1	&	0.15	&	0.2	&	0.25 \\
2& 0.35	&	0.45	&	0.5	&	0.6	&	0.7	&	0.8 \\
3& 0.01	&	0.3	&	0.55	&	0.65	&	0.8	&	0.95 \\
4& 0.04	&	0.06	&	0.08	&	0.1	&	0.3	&	0.6 \\
5& 0.05	&	0.6	&	0.8	&	0.9	&	0.95	&	0.99 \\
6& 0.01	&	0.05	&	0.1	&	0.6	&	0.7	&	0.9 \\
7& 0.01	&	0.03	&	0.07	&	0.1	&	0.15	&	0.75 \\
8& 0.29	&	0.31	&	0.33	&	0.35	&	0.37	&	0.39 \\
9& 0.25	&	0.27	&	0.29	&	0.31	&	0.33	&	0.35 \\
10& 0.21	&	0.23	&	0.25	&	0.27	&	0.29	&	0.31 \\
11& 0.05	&	0.2	&	0.27	&	0.33	&	0.39	&	0.45 \\
12& 0.05	&	0.1	&	0.2	&	0.3	&	0.4	&	0.4 \\
13& 0.3	&	0.35	&	0.4	&	0.45	&	0.5	&	0.55 \\
14& 0.15	&	0.18	&	0.21	&	0.24	&
0.27	&	0.3 \\ \hline
\end{tabular}
}
\end{center}
}
\end{table}

\nocite{Clertant:2017,cheung:2002,blanchard2011toxicity,berger2013statistical,blanchard2011toxicity,gezmu2006group,storer1989design,o1990continual,ji2007dose,ji2007bayesian,
ji2010modified,ji2013modified,yang2015integrated,guo2016bayesian,ivanova2007cumulative,
liu2015bayesian,cheung2011dose,yuan2016bayesian,goodman1995some,akaike1974new,schwarz1978estimating,
paoletti2004design,stylianou2002dose,babb1998cancer,jefferys1992ockham}

\end{document}